\documentclass[aps,prl,reprint,superscriptaddress]{revtex4-1}
\usepackage{subfiles}
\usepackage{graphicx}
\usepackage{textcomp}
\usepackage{gensymb}
\usepackage{xfrac}
\usepackage{amsmath}
\usepackage{amssymb}
\usepackage{wrapfig}
\usepackage[parfill]{parskip}
\usepackage{float}
\usepackage{textcomp}

\begin{document}

\title{Reciprocal Space Imaging of Ionic Correlations in Intercalation Compounds}
\author{Matthew J. Krogstad}
\author{Stephan Rosenkranz}
\affiliation{Materials Science Division, Argonne National Laboratory, Argonne,
IL 60439, USA}
\author{Justin M. Wozniak}
\affiliation{Data Science and Learning Division, Argonne National Laboratory, Argonne,
IL 60439, USA}
\author{Guy Jennings}
\affiliation{Advanced Photon Source, Argonne National Laboratory, Argonne,
IL 60439, USA}
\author{Jacob P. C. Ruff}
\affiliation{Cornell High Energy Synchrotron Source, Cornell University, Ithaca, NY 14853, USA}
\author{John T. Vaughey}
\affiliation{Chemical Sciences and Engineering Division, Argonne National Laboratory, Argonne,
IL 60439, USA}
\author{Raymond Osborn}
\affiliation{Materials Science Division, Argonne National Laboratory, Argonne,
IL 60439, USA}
\email{ROsborn@anl.gov}

\begin{abstract}
The intercalation of alkali ions into layered materials has played an essential role in battery technology since the development of the first lithium-ion electrodes. Coulomb repulsion between the intercalants leads to ordering of the intercalant sublattice, which hinders ionic diffusion and impacts battery performance. While conventional diffraction can identify the long-range order that can occur at discrete intercalant concentrations during the charging cycle, it cannot determine short-range order at other concentrations that also disrupt ionic mobility. In this article, we show that the use of real-space transforms of single crystal diffuse scattering, measured with high-energy synchrotron x-rays, allows a model-independent measurement of the temperature dependence of the length scale of ionic correlations along each of the crystallographic axes in a sodium-intercalated V$_2$O$_5$. The techniques described here provide a new way of probing the evolution of structural ordering in crystalline materials. 
\end{abstract}
 \date{\today}

\maketitle

\section{Introduction}
The cathodes of the first lithium-ion batteries developed over forty years ago were materials with van der Waals-bonded layers between which the lithium ions could be rapidly intercalated and de-intercalated \cite{Whittingham:1976dr,Winter:2018fz}. Since the first tests on Li$_x$TiS$_2$, many other intercalation compounds have been investigated as potential electrodes because of their rapid charging rates and inherent stability over multiple charge/discharge cycles \cite{Whittingham:2014dk}. Insertion of alkali ions between such weakly-bonded layers is highly reversible because the weak coupling of the intercalants to the host produces only minor structural modifications during the electrochemical cycle. The potential use of multivalent ions in batteries has renewed interest in such materials \cite{Tepavcevic:2015cl,SaiGautam:2015bg,Sun:2016co}, although weak coupling can compromise a battery's gravimetric and volumetric energy density. The transition metal oxides most frequently used as cathodes in consumer electronics, such as LiCoO$_2$ \cite{Mizushima:1980gf}, are also layered materials, albeit with stronger host-intercalant interactions \cite{Goodenough:2010wb}.

Even if the interactions of the alkali ions with the host lattice are weak, the Coulomb interactions between the intercalants cannot be ignored, especially at high concentrations \cite{Berlinsky:1979jf}. Voltage anomalies as a function of lithiation in TiS$_2$ were quickly ascribed to correlations between the intercalants induced by Coulomb repulsion \cite{Thompson:1978bz}, causing changes in the configurational entropy and consequently the chemical potential for ion insertion \cite{Reynier:2004cs}. Berlinsky showed that order-disorder transitions can be inferred from extrema in the voltage derivatives versus composition \cite{Berlinsky:1979jf}. Such anomalies have also been observed in transition metal oxide cathodes \cite{Reimers:1992fr,Li:1992fc}, where they have been modeled by both lattice gas \cite{Li:1992fc,Gao:1996cg,Kim:2001jg} and first-principles methods \cite{Wolverton:1998ek,VanderVen:1998ck,Ceder:1999bm}. Ordering of intercalants impacts battery performance by disrupting ionic mobility, and there is evidence that suppressing ordering can improve charging rates \cite{Wang:2018er}.

Order-disorder transitions are usually associated with discrete stoichiometries, corresponding to commensurate ordering of the intercalant sublattice. In Li$_x$CoO$_2$ and other transition metal oxides, ordering occurs at $x=\frac{1}{2}$ and a number of other rational fractions \cite{Reynier:2004cs,Toumar:2015hr,Chen:2017ji,Kaufman:2019dm}. Although it is possible to infer order-disorder transitions at these values from bulk measurements, direct evidence can only be obtained by the observation of (often weak) superlattice peaks from unit-cell expansion in diffraction measurements. The low cross section of lithium makes this difficult with x-ray scattering, unless associated changes in the host lattice space group are observable. For this reason, electron diffraction provided the first conclusive evidence of lithium-vacancy ordering in Li$_{0.5}$CoO$_2$ \cite{ShaoHorn:2003ev}, showing unit-cell doubling both within and perpendicular to the planes. However, conventional Bragg diffraction is only sensitive to long-range order and cannot be used to probe short-range correlations that will also play a role in decreasing ionic mobility across the whole composition range. Understanding such correlations is important in modeling diffusion kinetics, for example as a function of concentration fluctuations close to the electrolyte interface.

The most powerful technique to probe short-range correlations is single crystal diffuse scattering, using either x-rays or neutrons, which comprises all the scattering that results from nanoscale deviations from the average crystalline structure \cite{Welberry:1995wm,Frey:1995tp,Nield:2001wg}. The average occupancies of all the atomic sites within the translationally invariant unit cell of a crystal can be determined from Bragg peak intensities using standard crystallographic techniques, but, apart from thermal diffuse scattering from lattice vibrations, all the scattering between (and under) the Bragg peaks results from the presence of local disorder, often in the form of point defects, such as interstitials and vacancies, or of short-range order from defect-defect correlations. When measured over a sufficiently large volume of reciprocal space, diffuse scattering contains a true thermodynamic average of disorder scattering from the entire sample and so is complementary to techniques that utilize x-ray coherence to reconstruct real-space images of atomic correlations over a limited coherence volume.

Although it is an extremely powerful technique, single crystal diffuse scattering has not been as widely utilized as, for example, pair-distribution-function (PDF) measurements of polycrystalline materials \cite{Egami:2003hf}, because of the challenge of both measuring and modeling large volumes of reciprocal space (\textbf{Q}-space). By contrast, the PDF technique, which gives a one-dimensional spherical average of interatomic vector probabilities, has become a standard tool in materials science because the information can be interpreted intuitively and modeled effectively with existing software. In recent years, Thomas Weber and colleagues have shown that it is possible to extend the PDF technique to three dimensions by transforming single crystal diffuse scattering into real space \cite{Weber:2012en}. The result is a Patterson function, containing the summed probabilities of interatomic vectors in the crystal with contributions from both the average structure and local disorder. By excluding Bragg scattering from the real space transform, the resulting PDF function only includes those interatomic vectors whose probabilities differ from the average (3D-$\Delta$PDF). This is not possible in polycrystalline data because of the substantial overlap of Bragg and diffuse scattering intensity, but in single crystals, it allows a determination of the length scales of short-range order extending over 100~\AA\ or more. By exploiting a new generation of fast area detectors, it is possible to measure sufficiently complete 3D volumes of reciprocal space in under 20 minutes using high-energy x-rays at a synchrotron source, enabling the generation of robust $\Delta$PDFs fast enough to allow detailed studies as a function of temperature.

We have applied this technique  to measure ionic correlations in single crystals of V$_2$O$_5$ intercalated with sodium ions. In $\beta$-Na$_x$V$_2$O$_5$, the sodium ions are confined to two-leg ladders \cite{Wadsley:1955jg}, but the configuration of ions within the partially-occupied ladders cannot be determined by powder diffraction alone. The $\Delta$PDFs reveal both the tendency for sodium ions to form zig-zag configurations within each ladder, which are in phase with those in neighbouring ladders, as well as the temperature dependence of the correlation length of this ordering in all three crystallographic directions, obtained from the exponential decay of the $\Delta$PDF intensities. It has previously been reported that this material undergoes a structural phase transition above 200~K \cite{Kanai:1982fx}, but our results show that this temperature marks a crossover from two- to three-dimensional correlations without true long-range order being established down to 30~K. In the following article, we describe the measurement process in more detail, the steps used to generate the 3D-$\Delta$PDF results, and the subsequent analysis to extract ionic correlation lengths, before discussing the implications of this technique for studying battery electrodes and other disordered materials.

\begin{figure}[!b]
\centering
\includegraphics[keepaspectratio,width=\columnwidth]{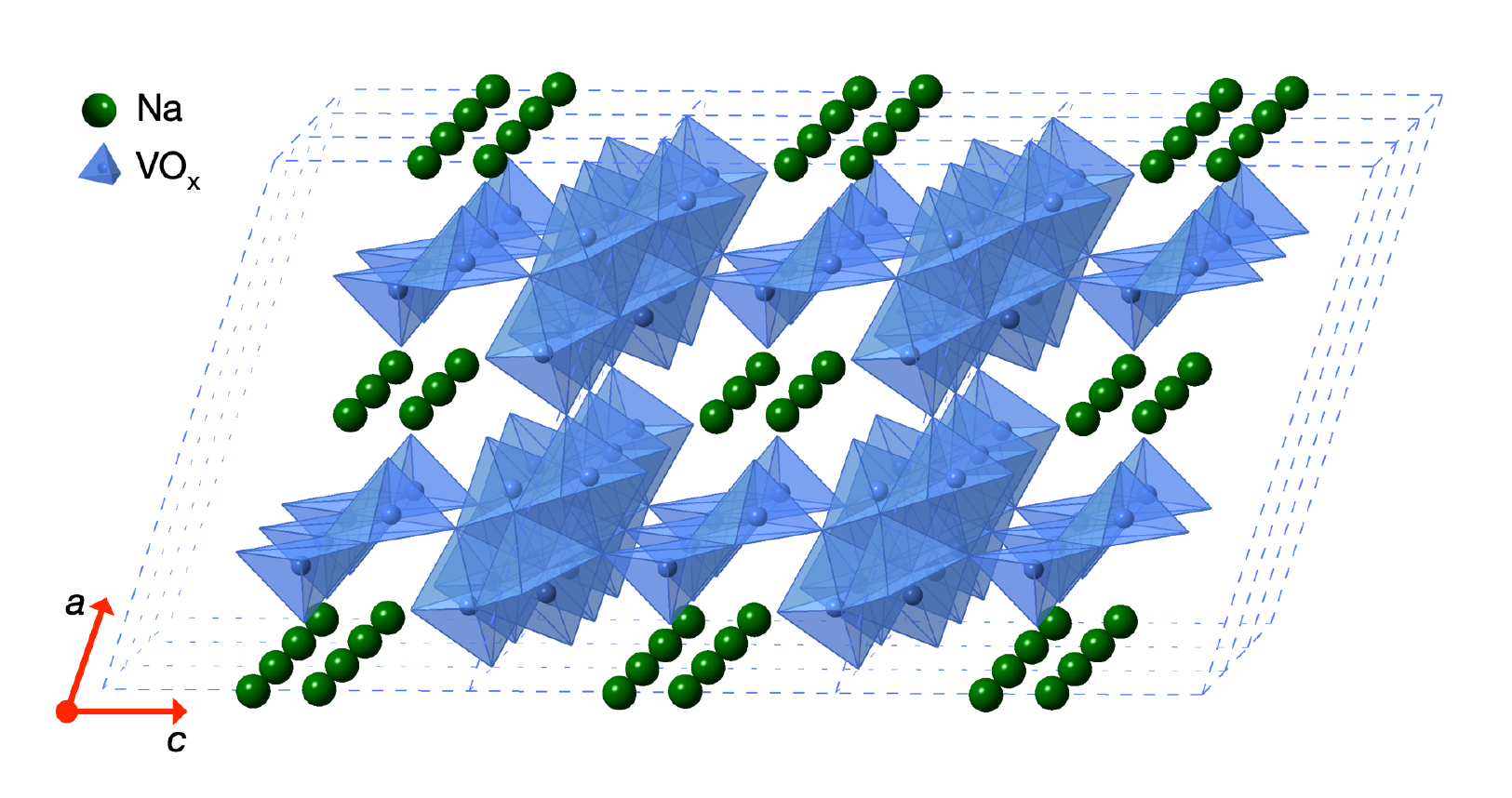}
\caption{The monoclinic structure of Na\textsubscript{0.45}V\textsubscript{2}O\textsubscript{5}. The sodium ions (green) sit on two-leg ladders with rungs parallel to the crystallographic $c$-axis in channels parallel to the $b$-axis formed by the lattice of edge- and corner-shared vanadium oxide octahedra and square pyramids (blue). Neigbouring ladders along the $a$-axis, \textit{i.e.}, at $\pm\frac{1}{2}a$, are displaced by $\frac{1}{2}b$.  Dashed lines show the unit cell boundaries.}
\label{FIg1}
\end{figure}

\begin{figure*}[!t]
\centering
\includegraphics[keepaspectratio,width=2\columnwidth]{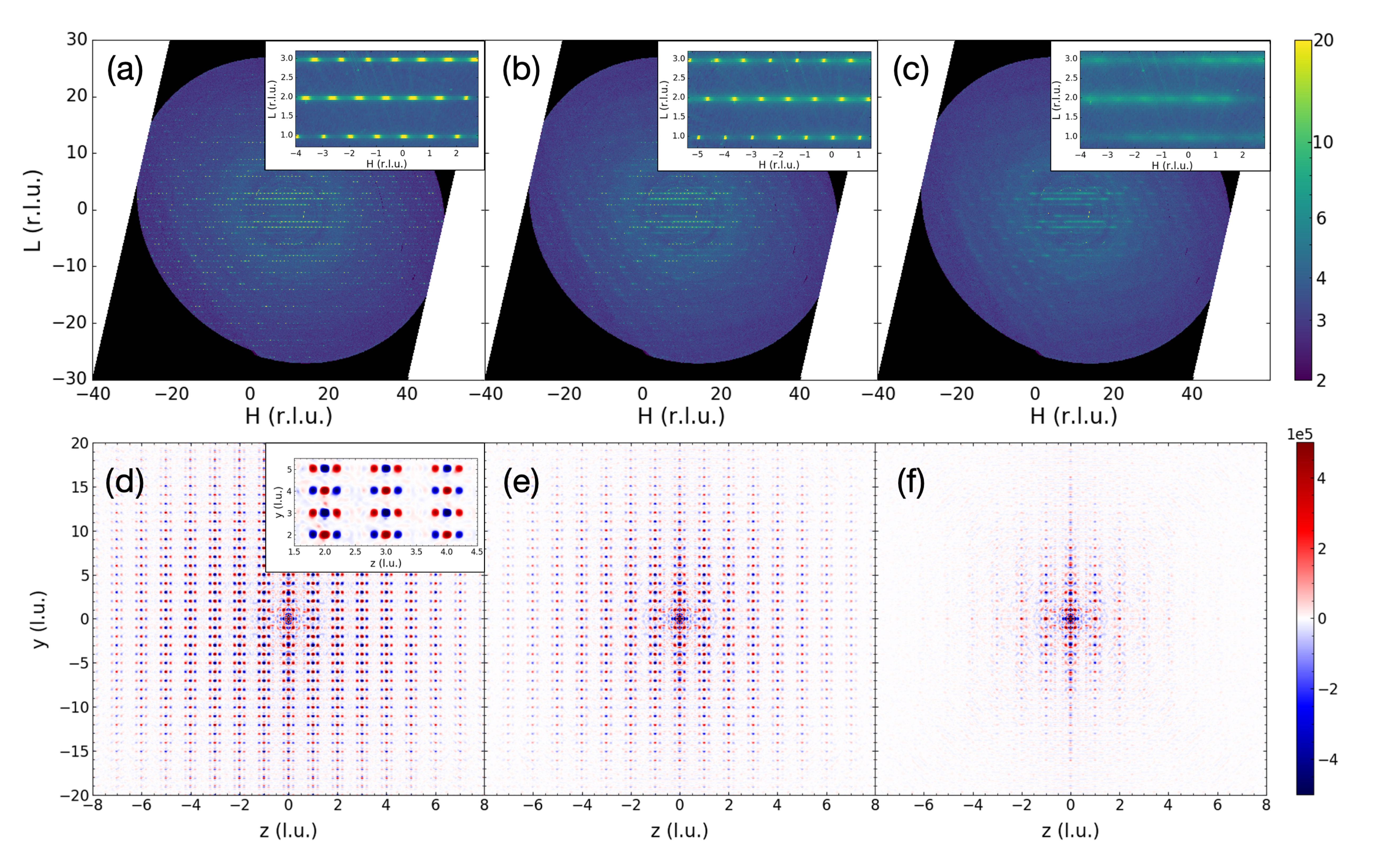}
\caption{Diffuse scattering from Na\textsubscript{0.45}V\textsubscript{2}O\textsubscript{5} in the \textbf{Q} =  ($H$$\frac{1}{2}$$L$) plane at (a) 50~K, (b) 200~K, and (c) 250~K, displayed in reciprocal lattice units (r.l.u.), and the resulting $\Delta$PDF in the (0yz) plane at (d) 50~K, (e) 200~K, and (f) 250~K, displayed in lattice units (l.u. with $a=15.36$~\AA, $b=3.61$~\AA, $c=10.05$~\AA\ at 100~K)}
\label{Fig2}
\end{figure*}

\section{Results}

The orthorhombic structure of pure V\textsubscript{2}O\textsubscript{5} consists of both edge- and corner-shared square pyramids that form van der Waals-bonded planes, between which a variety of cations may be intercalated \cite{Marley:2015hm}. However, for some intercalants above a critical concentration, the structure transforms to a monoclinic $\beta$-phase (space group $C2/m)$ with a more complex network of vanadium oxide octahedra and square pyramids that contain one-dimensional channels, into which the intercalated cations are inserted (Fig. 1) \cite{Wadsley:1955jg,Hiroyuki:2013kn}. In $\beta$-Na\textsubscript{x}V\textsubscript{2}O\textsubscript{5} ($0.2\lesssim x\lesssim0.5$), the intercalants partially occupy a sublattice of two-leg ladders, whose legs are parallel to the crystal's \emph{b}-axis and whose rungs are parallel to the \emph{c}-axis. The ladder sites would be filled with a value of $x=\frac{2}{3}$ , but stereochemical constraints should prevent occupations greater than 50\% since the ladder rungs are too short ($<$2~\AA). It has been postulated that the excess sodium ions would occupy additional octahedral and tetrahedral sites for $x>0.33$ in a $\beta^\prime$-phase \cite{Galy:1970dy}, but these have never been conclusively observed. In the Supplementary Information, we discuss features in our data that provide evidence in favour of the octahedral site occupation preferred in Ref. \citenum{Galy:1970dy}.

We measured diffuse x-ray scattering on single crystals of $\beta^\prime$-Na\textsubscript{x}V\textsubscript{2}O\textsubscript{5} with \emph{x} = 0.45, corresponding to 68\% occupation, using high-energy monochromatic beams at two synchrotron x-ray sources, the Advanced Photon Source (APS) and the Cornell High Energy Synchrotron Source (CHESS). Advances in the dynamic range and speed of x-ray area detectors now allow both the Bragg peaks and diffuse scattering to be measured efficiently, with three-dimensional volumes of scattering in reciprocal space, S(\textbf{Q}), collected in under 20 minutes at each temperature. More details are given in the Methods section and the Supplementary Information.

The diffuse scattering is mostly confined to rods that are parallel to $H$, \textit{i.e.}, orthogonal to the ladders, occurring at half-integer values of $K$ (Fig. 2), from which we can infer that there is predominantly two-dimensional short-range order within the ladder planes generated by ionic correlations within the sodium sublattice that tend to double the unit cell along \textit{b}. A sinusoidal modulation of the rod intensities as a function of $L$, \emph{i.e.}, parallel to the direction of the ladder rungs, whose periodicity is the inverse of the rung length, indicates that site occupations across each ladder rung are strongly correlated. At room temperature, there are no sharp peaks along the rods (Fig. 2c), which are broad laterally ($\sim$0.1~\AA\textsuperscript{-1} along $K$ and $\sim$0.13~\AA\textsuperscript{-1} along $L$), indicating that correlation lengths are of the order of 10-20~\AA. However, below $\sim$230~K, peaks start to appear at all integer values of $H$ (Fig. 2b), growing steadily in intensity down to the lowest measured temperature of 30~K (Fig. 2a). These peaks are evidence of the development of longer-range three-dimensional correlations within the sodium sublattice below $\sim$230~K, in agreement with previous x-ray measurements \cite{Kanai:1982fx,Yamaura:2002ia}.

\begin{figure*}[!tp]
\centering
\includegraphics[keepaspectratio,width=2\columnwidth]{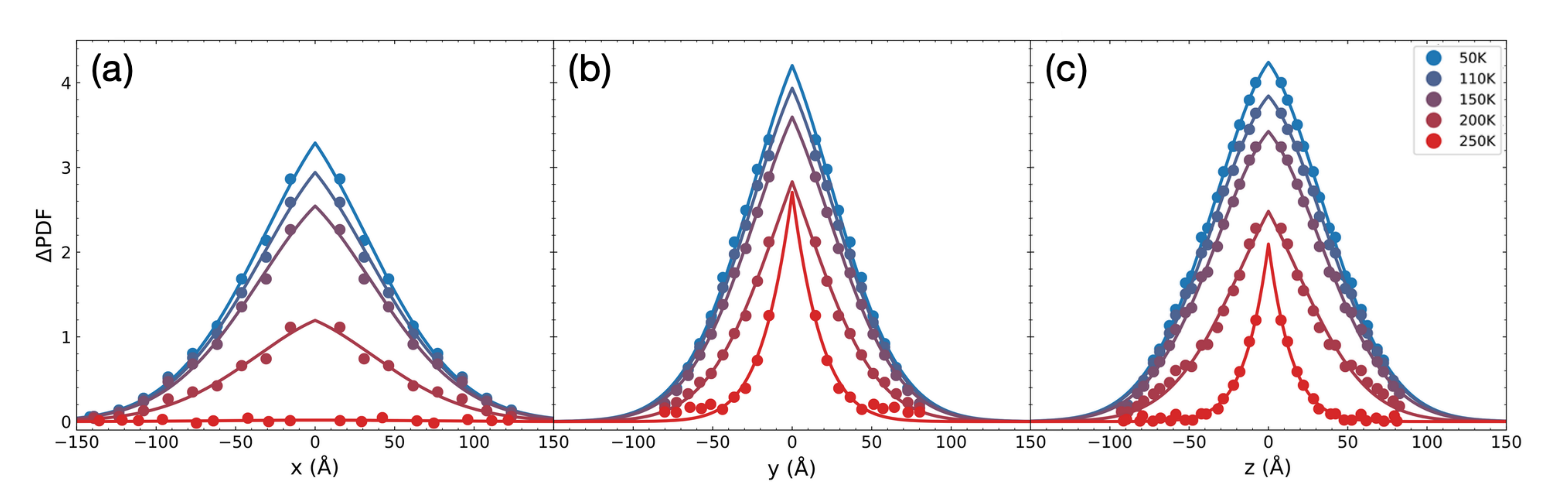}
\caption{Fits to the temperature dependence of the $\Delta$PDF peak intensities along the (a) $x$-direction, (b) $y$-direction, and (c) $z$-direction. The model is the product of an exponential decay from the pair correlations and a Gaussian envelope from the finite \textbf{Q}-resolution. The data are shown from 50~K (blue) to 250~K (red) in steps of 50~K.}
\label{FIg3}
\end{figure*}

Most analyses of diffuse scattering require a calculation of the \textbf{Q}-variation of the data using atomistic models that parametrize the disorder, but an alternative approach is to Fourier transform the data to generate real-space PDFs. This has the advantage of converting complicated intensity distributions in reciprocal space into discrete peaks in real space, whose positions and intensities are given by the interatomic vectors present in the disordered structure and their weighted probabilities, respectively. A further simplification is to include in the generated PDFs only those peaks whose probabilities deviate from the average structure. This is possible using the ``Punch and Fill'' method pioneered by Weber \emph{et al} \cite{Weber:2012en}, which utilizes the fact that the total scattering can be separated into two components: one representing the average crystal, \textit{i.e.}, the Bragg peaks, and the other representing scattering from defects.
\begin{equation}
I_{tot}(\boldsymbol{\mathrm{Q}})=F(\boldsymbol{\mathrm{Q}})F^{\ast}(\boldsymbol{\mathrm{Q}})=|F_{H\! K\! L}(\boldsymbol{\mathrm{Q}})|^2+|\Delta F(\boldsymbol{\mathrm{Q}})|^2
\end{equation}
where $I_{tot}(\boldsymbol{\mathrm{Q}})$ is the measured total scattering and $F(\boldsymbol{\mathrm{Q}})$ is proportional to the Fourier transform of the electron density. The Fourier transform of this scattering function is also separable.
\begin{equation}
\begin{split}
P_{tot}(\boldsymbol{\mathrm{r}})&=\mathrm{FT}[|F_{H\! K\! L}(\boldsymbol{\mathrm{Q}})|^2]+\mathrm{FT}[|\Delta F(\boldsymbol{\mathrm{Q}})|^2] \\
&=P_{H\! K\! L}(\boldsymbol{\mathrm{r}})+\Delta P(\boldsymbol{\mathrm{r}})
\end{split}
\end{equation}
$P_{H\! K\! L}$ is the Patterson function of the average structure while $\Delta P(\boldsymbol{\mathrm{r}})$, or the 3D-$\Delta$PDF, is the difference Patterson function due to the disorder. The ``Punch and Fill'' method isolates $\Delta P(\boldsymbol{\mathrm{r}})$ by eliminating the Bragg peaks, \textit{i.e.}, by removing the scattering in a small sphere around each Bragg peak and interpolating over the missing data, before performing the Fourier transform. The resulting $\Delta$PDF contains both positive and negative values for interatomic vectors that are, respectively, more or less probable than in the average structure.

We have used this technique to produce $\Delta$PDFs, which eliminate any contribution from the framework V$_2$O$_5$ lattice to first order and leave only the sodium-sodium pair correlations. In the average structure determined by powder diffraction \cite{Hughes:1983ux}, the sodium sites are randomly occupied but our data shows that there are significant short-range correlations. In Fig. 2d-f, the $\Delta$PDF in the $b-c$ plane is displayed as a symmetric log plot consisting of triplet patterns of red and blue dots corresponding to positive and negative probabilities, respectively (see the inset to Fig. 2d). These triplet motifs are consistent with the allowed sodium-sodium interatomic vectors; the two-leg ladders of the real-space structure produce three-leg ladders in the PDFs, since they include interatomic vectors that connect sites on the left leg to sites on the right leg and \textit{vice versa}. 

There are weaker $\Delta$PDF peaks in other planes that could either be associated with additional sodium sites, such as those proposed by Galy \textit{et al} \cite{Galy:1970dy}, or with relaxations of the vanadium and oxygen ions. In the scattering data, additional superlattice peaks are also observed at $K=\pm\frac{1}{6}$, below $\sim$130~K, which is the temperature of a metal-insulator transition ascribed to V$^{4+}$/V$^{5+}$ charge ordering  \cite{Yamada:1999kn, Yamaura:2002ia}. These give rise to a weak modulation of  the $\Delta$PDF intensity along the $y$-axis, but does not appear to affect the sodium correlations significantly. These additional features are discussed in the Supplementary Information.

Some immediate conclusions can be drawn by inspection of the $\Delta$PDF data. Figure 2 shows that the nearest-neighbour sites within the ladder are less likely to be occupied, but next-nearest-neighbour sites are more likely to be occupied, leading to a zig-zag configuration of occupied sites. This is expected from considerations of both stereochemistry and Coulomb repulsion. Furthermore, these zig-zag configurations are in phase with neighbouring ladders, since the vectors connecting sites on the same leg (left or right), \textit{i.e.}, with $\pm ma \pm nc$, where $m$ and $n$ are integers, have positive probability whereas those connecting the opposite legs are negative. At 250~K (Fig. 2f), these correlations extend over only a few neigbouring sodium ladders, but as the temperature is lowered, the correlation length increases significantly, as seen at 200~K (Fig. 2e). At 50~K, the correlations extend over more than 100~\AA\ in all directions (Fig. 2d). Since we have not removed the low-temperature peaks from the $k=\frac{1}{2}$ rods, it is evident that the $\Delta$PDFs are able to track continuously the growth in correlations from above to below the transition temperature, with the $\Delta$PDF peak intensities providing a measure of the incipient order parameter.

\begin{figure}[!bp]
\centering
\includegraphics[keepaspectratio,width=\columnwidth]{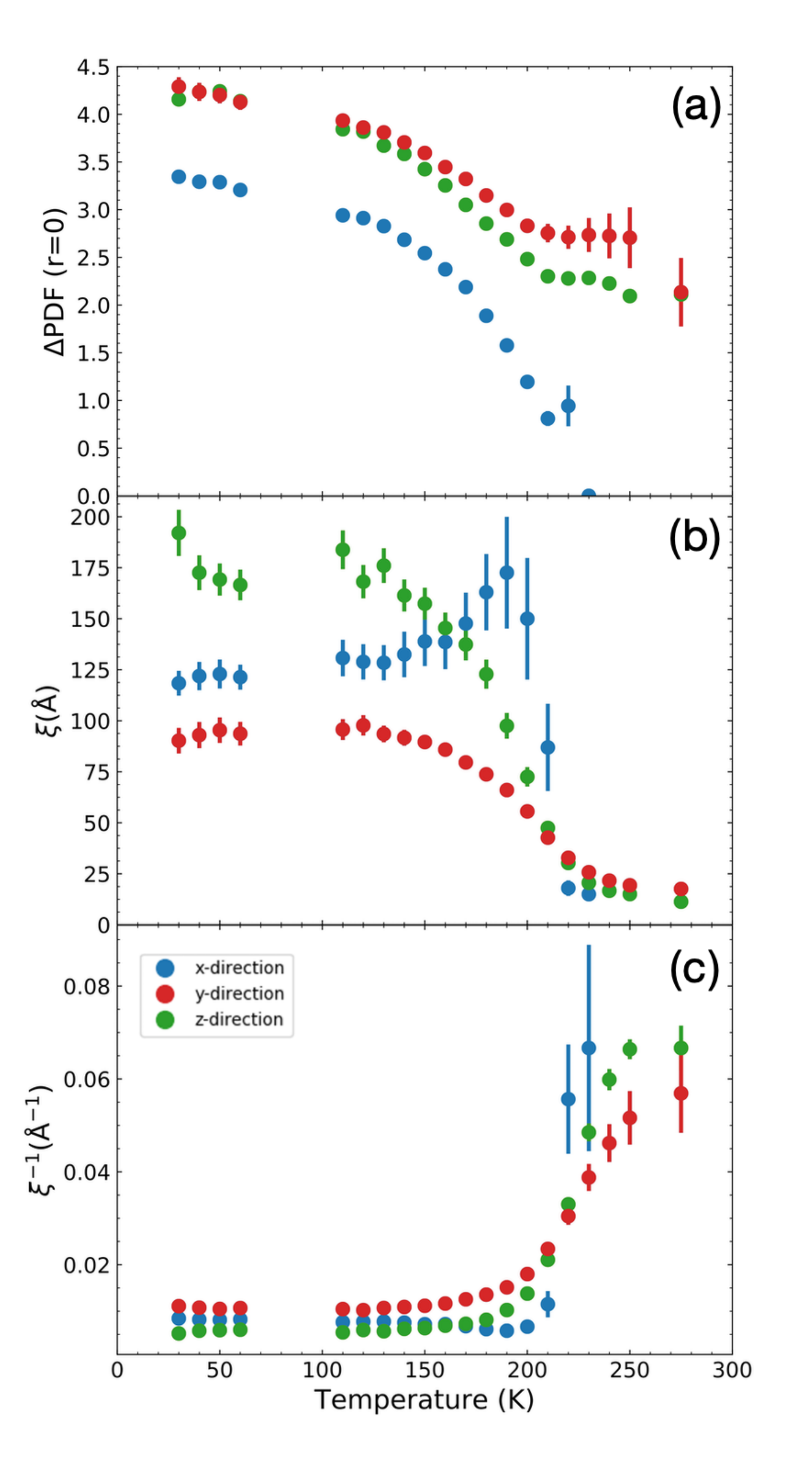}
\caption{The results of fitting the $\Delta$PDF peak intensities to a decaying exponential as a function of temperature along the three crystallographic axes: (a) the amplitude of the $\Delta$PDF exponential decays at $r=0$ along the $x$ (blue), $y$ (red), and $z$ (green) directions, (b) the correlation length of sodium pair correlations derived from the decay constant, and (c) the inverse correlation length. A beam failure prevented measurements between 60 and 110~K. Reliable fits along the $x$-direction were not possible above 230~K because the $\Delta$PDF intensities were too weak.}
\label{FIg4}
\end{figure}

To quantify the growth in correlations, we have modeled the dependence of the $\Delta$PDF peak intensities on interatomic distance. Figure 3 shows that these intensities follow an exponential decay at high temperature, consistent with a one-dimensional Ornstein-Zernike function \cite{Collins:1989th}. At lower temperature, it is necessary to include the effect of the finite resolution in reciprocal space, which produces in an envelope function in real space. Its Gaussian width can be estimated by fitting the total PDF, \textit{i.e.}, the data transformed without eliminating the Bragg peaks, which represents long-range crystalline order. The $\Delta$PDF intensities have then been fit to the product of this Gaussian envelope function with a temperature-dependent exponential decay, whose decay constant is a direct measure of the correlation length, $\xi$. More details are provided in the Supplementary Information. 

The temperature dependence of these fits is shown in Fig. 3 along all three crystallographic directions. As the correlation length approaches 200~\AA, the effect of the finite resolution leads to an increase in the the error bars although $\xi$ and the $r=0$ amplitude of the $\Delta$PDF correlations is measurable down to 30~K.

\section{Discussion}
The elimination of the interatomic vector probabilities of the average structure from the PDF allows the range of short-range order to be determined over length scales that exceed 100~\AA, which means that the correlation of ions undergoing an order-disorder transition can be monitored as a function of temperature from well-above to well-below the transition (Fig. 4). In $\beta$-Na\textsubscript{0.45}V\textsubscript{2}O\textsubscript{5}, there is a gradual increase in the correlation lengths along the $y$ and $z$ directions, $\xi_y$ and $\xi_z$, from 275~K to 230~K, which is easier to see in plots of $\xi^{-1}$ \textit{vs} temperature (Fig. 4c). At $\sim$230~K, this induces a sharp increase in correlation length of the sodium pair correlations along the $x$-axis, $\xi_x$, \textit{i.e.}, perpendicular to the sodium ladder planes, marking a cross-over from two- to three-dimensional correlations (Fig. 4b). However, it does not appear to be a true second-order phase transition, since $\xi_x$ saturates at 190~K with a value of only 170~\AA, corresponding to approximately 10 unit cells or 20 ladder planes. The increase in $\xi_x$ in turn generates a more rapid increase in $\xi_y$ and $\xi_z$, again without leading to a true divergence. The maximum correlation length at 30~K is still under 200~\AA. Instead of a discontinuity in the slope of the inverse correlation lengths, $\xi_y^{-1}$ and $\xi_z^{-1}$, defining a unique transition temperature, there is a crossover between two temperature regimes at $\sim$200~K close to where $\xi_x$ saturates (Fig. 4c). 

We propose that the saturation in $\xi_x$ is a consequence of frustration.  Due to the body-centring of the monoclinic structure, the ladder sites in the $y$-$z$ planes at $x=0$ and $x=\frac{1}{2}a$ are shifted by $\pm\frac{1}{2}b$, which makes the strength of the Coulomb interactions independent of the phase of the zig-zag site occupations between such neighbouring planes. This degeneracy implies that the phase must be stabilized by next-nearest-neighbour interactions, which may be weak enough to allow stacking faults along the $x$-axis, preventing true long-range order. Disorder of the excess sodium ions that cannot be accommodated on the ladder sites would contribute to this frustration providing local pinning of the zig-zag phase. The fall in $\xi_x$ below 190~K indicates that any such frustration mechanism is increased by the growth in intraplanar correlations. This could be tested by further measurements on $\beta$-Na\textsubscript{0.33}V\textsubscript{2}O\textsubscript{5}, corresponding to precisely 50\% occupation of the ladder sites.

The intensities of the $\Delta$PDF peaks increase monotonically with decreasing temperature (Fig. 4a). Along the $x$-axis, the temperature dependence of the amplitude resembles an incipient order parameter, with a value close to 0 above the transition (see Fig. S5 in the Supplementary Information). However, along the $y$ and $z$ directions, there is a much weaker temperature dependence reflecting the strong nearest-neighbour correlations at all temperatures. 

\section{Conclusion}
Transforming x-ray diffuse scattering data into 3D-$\Delta$PDFs produces a remarkable simplification in how the data are represented. It is possible to interpret the short-range order in the crystal structure without a detailed simulation of the disorder, which has in the past required the optimization of a large number of parameters over a substantial volume of reciprocal space. In fact, the PDF intensities directly determined from the transformed data are related by simple analytic functions to the Warren-Cowley parameters that are frequently used to parametrize diffuse scattering models \cite{Welberry:1995wm}. Even without a well-defined model of the underlying disorder, our results show that the spatial dependence of the PDF intensities provides a method of extracting correlation lengths as a function of temperature or other parametric variable.

The ability to generate a real-space ``image'' from reciprocal space data makes this a powerful tool in the investigation of intercalation compounds, as it is especially suited to the measurement of ionic correlations on a sublattice that is distinct from the host structure. Although we have not yet tested this technique on lithiated compounds, we believe that the elimination of the average structure of heavier elements will allow lithium-lithium correlations to be measured in spite of their low cross section.  In this article, the method has allowed us to show that the apparent order-disorder transition reported in Na$_x$V$_2$O$_5$ is actually a crossover from two-dimensional to three-dimensional correlations, with interplanar correlations saturating at only $\sim$150~\AA, probably because of frustration. Such detailed insight into intercalant concentrations is not possible by any other method.

Although we have focused on layered materials, the technique can be utilized in more three-dimensional insertion compounds such as the oxide spinels \cite{Kim:2001jg}, as well as many other disordered materials. The principal limitation of this technique is the need for single crystals, which makes \textit{in situ} experiments within functioning battery cells challenging, although not impossible. Sample thicknesses can be 100~$\mu$m or less, and provided the rest of the cell is polycrystalline, the real-space transform will only contain discrete peaks that correspond to the crystalline electrode. Measuring the phase diagram of order-disorder transitions as a function of both temperature and intercalant concentration will allow the strength of inter-ionic interactions to be determined and improve our understanding of the how they limit ionic mobility. Such measurements could also be used to evaluate the effectiveness of strategies for mitigating these limitations, for example by co-intercalation with aliovalent cations  \cite{Meethong:2009bj} or substitution of transition metals on the vanadium sites \cite{Jovanovic:2018vj}, in order to disrupt ordering phenomena. 

\section{Methods}
\subsection{Synthesis}
Na$_x$V$_2$O$_5$ crystals were grown using a self-flux technique.  Five grams of vanadium pentoxide (V$_2$O$_5$, Aldrich, $>98$\%) was placed into a nickel crucible and placed in an oven inside an Ar-atmosphere glovebox.  The sample was heated to 720\textdegree C.  While liquid, the crucible was removed from the furnace and a stoichiometric amount of sodium iodide (NaI, Aldrich 99.5\%) was added to the liquid.  The sample was then put back in the furnace and cooled to 650\textdegree C over a one hour period, then radiatively cooled to room temperature.  Single crystals were isolated from the flux. EDX analysis showed that the sodium concentration, when normalized to the nominal vanadium stoichiometry, was  $x=0.45\pm0.02$. 

\subsection{X-ray Scattering}
Three-dimensional volumes of diffuse x-ray scattering were collected at the Advanced Photon Source (APS) and the Cornell High Energy Synchrotron Source (CHESS). The APS data was measured on Sector 6-ID-D using an incident energy of 87.1~keV and a Dectris Pilatus 2M with a 1~mm-thick CdTe sensor layer. The CHESS data were measured on beamline A2 using an incident beam energy of 27.3~keV and a Dectris Pilatus 6M detector with a 1~mm-thick Si sensor layer. The data were collected from 30~K to 300~K, with samples cooled by flowing He gas below 100~K and N2 gas above 100~K. During the measurements, the samples were continuously rotated about an axis perpendicular to the beam at 1\textdegree\ per second over 370\textdegree, with images read out every 0.1~s. Three sets of rotation images were collected for each sample at each temperature to fill in gaps between the detector chips. The resulting images were stacked into a three-dimensional array, oriented using an automated peak search algorithm, and transformed in reciprocal space coordinates using the software package CCTW (Crystal Coordinate Transformation Workflow) \cite{Jennings:CCTW}, allowing S(\textbf{Q}) to be determined over a range of $\sim$$\pm $15~\AA$^{-1}$ in all directions. Further details are given in the Supplementary Information.

\section{End Notes}
\subsection{Acknowledgements}
This work was supported by the U.S. Department of Energy, Office of Science, Materials Sciences and Engineering Division and Scientific User Facilities Division. X-ray experiments were performed at the Advanced Photon Source, which is supported by the Office of Basic Energy Sciences under Contract No. DE-AC02-06CH11357 and the Cornell High Energy Synchrotron Source (CHESS), which is supported by the NSF and NIH/NIGMS via NSF award DMR-1332208. We thank Douglas Robinson and Xiaoyi Zhang for technical support during the experiments, Alexander Rettie for performing the EDX analysis, Thomas Weber and Arkadiy Simonov for discussions about the $\Delta$PDF technique, Branton Campbell for help with the formalism of transforming the data to reciprocal space, and Peter Zapol and Charlotte Haley for discussions about interpreting the results. Crystal structure images were generated using CrystalMaker\textsuperscript{\tiny\textregistered}, CrystalMaker Software Ltd, http://www.crystalmaker.com.

\subsection{Author contributions} 
Samples were prepared by J.T.V. and prepared for measurement by M.K. The experiments were devised by M.K., S.R., and R.O. The x-ray experiments  were performed by M.K., S.R., J. R., J.M.W., and R.O. The data were analyzed by M.K., R.O., J.M.W., and G.J., using software written by G.J., M.K., R.O., and J.M.W. The manuscript and supplementary information were written by R.O. with input from all the authors.

\begin{figure}[!h]
\includegraphics[width=\columnwidth]{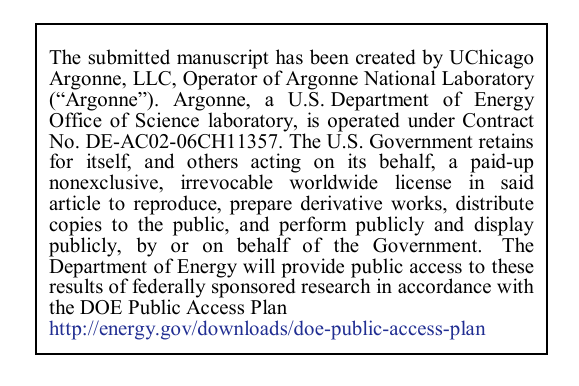}
\end{figure}
\pagebreak

\section{Experimental Method}

\subsection{Data Reduction}
The diffuse scattering data were measured using monochromatic beams of high-energy x-rays at the Advanced Photon Source (APS) on Sector 6-ID-D and at the Cornell High Energy Synchrotron Source (CHESS) on Beamline A2, with incident energies of 87.1~keV and 27.3~keV respectively. The data were collected on area detectors in the Dectris Pilatus series, which have 20 bit dynamic range, low backgrounds, and fast readout times. The APS data were collected on a Pilatus 2M detector with a 1~mm CdTe sensor layer optimized for high x-ray energies, and the CHESS data were collected on a Pilatus 6M detector with a 1~mm silicon sensor layer.  The distance of the detectors behind the sample were 650~mm at the APS and 700~mm at CHESS.  

During data collection, the sample was continuously rotated in the x-ray beam through more than 360\textdegree\ while an area detector captured images every 0.1\textdegree. This rotation method is in common use in crystallography, where the image frames are used to determine Bragg peak positions and intensities before the rest of the data is discarded. However, for diffuse scattering, all the data is preserved and transformed into three-dimensional reciprocal space coordinates, \textbf{Q}=(\textit{HKL}). 

\begin{figure}[!b]
\includegraphics[width = \columnwidth]{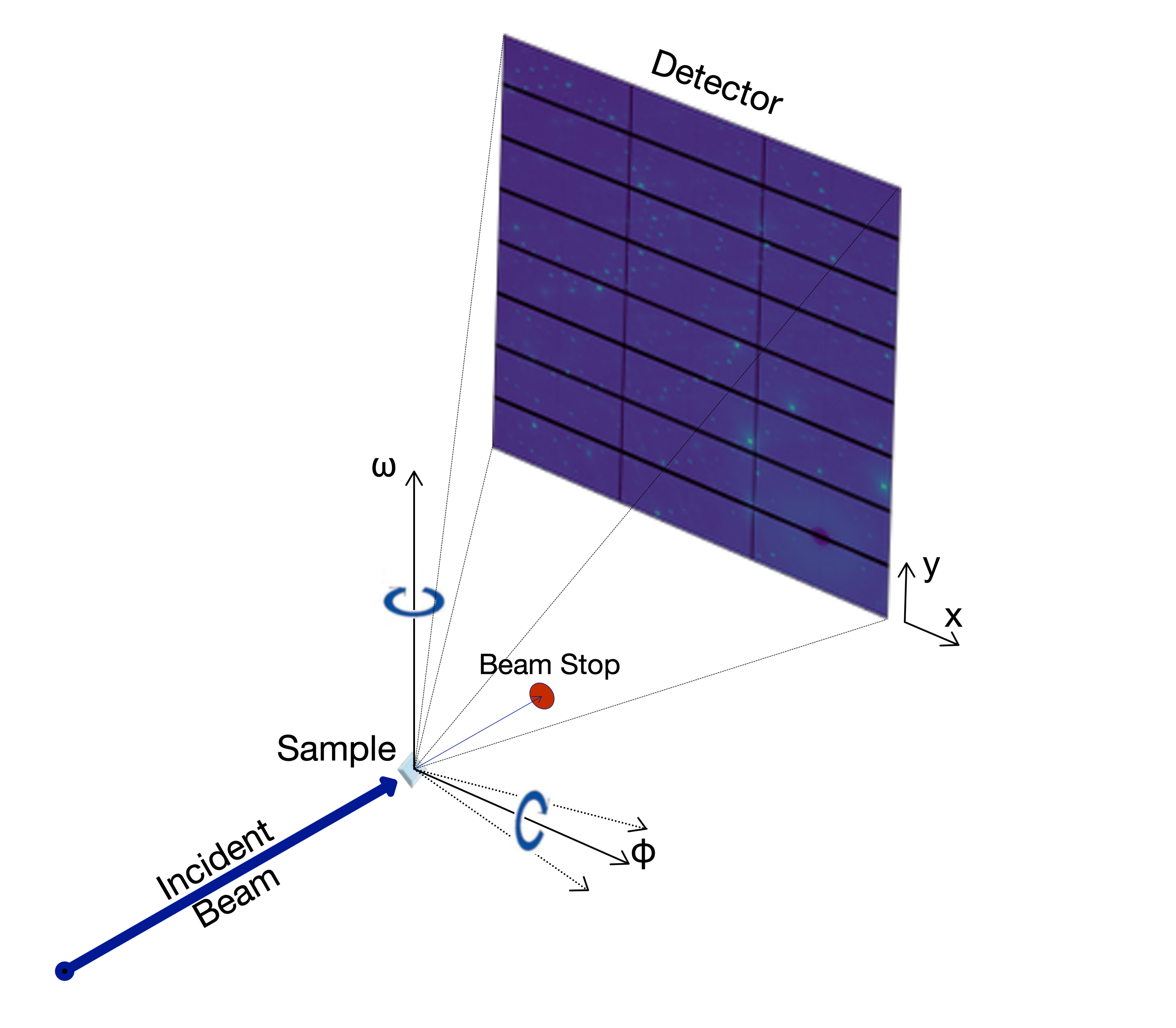}
\caption{Experimental configuration on Sector 6-ID-D at the Advanced Photon Source. The detector is mounted perpendicular to the incident beam 650~mm behind the sample. For each measurement, the sample is rotated continuously through 370\textdegree\ about the $\phi$ axis, which is in the horizontal plane. The rotation is repeated twice, with the $\phi$ axis rotated about the vertical $\omega$ axis by $\pm15$\textdegree\ ($\omega=0$ corresponds to the $\phi$ axis perpendicular to the beam) and the detector translated by 5~mm in both the $x$ and $y$ directions perpendicular to the beam. 
\label{FigureS1}}
\end{figure}

Here is an outline of the experimental workflow. Further details will be given in another publication. 

\begin{enumerate}
\item Before mounting the crystal, the detector distance, as well as yaw and pitch corrections, were determined by measurement of a CeO$_2$ powder standard, analyzed using the PyFAI package \cite{Ashiotis:2015ez}.

\item The sample was mounted on a goniometer with a $\phi$ motor allowing continuous 370\textdegree\ rotations at 1\textdegree\ s$^{-1}$ about a horizontal axis perpendicular to the beam, during which frames were captured  at 10~Hz on the detector, \textit{i.e.}, every 0.1\textdegree. After a complete rotation, the detector was twice moved on a translation stage by 5~mm in both the horizontal and vertical directions perpendicular to the beam and the rotation was repeated each time.  This ensured that scattering into the gaps between the detector chips in one rotation scan were measured in one of the other two scans. At the APS, the axis of the $\phi$ rotation was also shifted by $\pm15$\textdegree\ about the vertical direction perpendicular to the $\phi$ axis, \textit{i.e.}, the $\omega$ axis in Fig. \ref{FigureS1}. This is because we observed detector artifacts caused by Compton scattering within the sensor layer when intense Bragg peaks hit the detector. This caused a disc of excess counts around each Bragg peak occurring in a few frames where the Bragg intensity was strong. By rotating about the $\omega$ axis between each rotation scan, these discs were also rotated in reciprocal space, allowing the spurious counts to be masked at one $\omega$ setting while ensuring that those regions of reciprocal space were covered by the other scans. Three sample rotations took 20 minutes to complete. Nearly all the data shown in this article is from the later APS experiment, but the analysis was initially developed using the CHESS data, which produced very similar results and was used in the order parameter analysis shown below.

At the APS the sample temperature was controlled by a cryostream provided by an Oxford Instrument n-HeliX Cryodrive, with a helium gas flow for temperatures from 30~K to 100~K and nitrogen gas flow for temperatures above 100~K. At CHESS, a nitrogen cryostream was used with a base temperature of 100~K. Use of cryostreams ensured that the only background arose from air scattering from the incident beam before the beam stop. This was estimated by collecting data without a sample in the beam. Correcting for this background is discussed in the next section.

\item After the sample rotations, the measured frames collected as TIFF or CBF images were stacked into three-dimensional HDF5 arrays, which were then externally linked to files containing the experimental metadata stored using the NeXus format \cite{Koennecke:2006be}. At the APS, the three arrays (one for each $\phi$ rotation) were $\sim$30~GB in size uncompressed.

\item The locations of all the Bragg peaks in the 3D data volume were determined by an automated search followed by a refinement of the UB matrix using the Python package NXrefine \cite{Osborn:NXrefine}.

\item The UB matrix was used to transform separately the three HDF5 arrays into reciprocal lattice coordinates and then merged using the CCTW package \cite{Jennings:CCTW}. The transformed data were stored in separate HDF5 files that were also externally linked to the same NeXus file as the raw data.

\item The resulting data were analyzed using the NeXpy package \cite{Osborn:NeXpy}, a PyQt GUI, which allows the data to be visualized in Matplotlib windows \cite{Hunter:2007} and manipulated in an IPython shell \cite{Perez:2007}. The Punch-and-Fill method described in the next section was implemented as Python scripts within NeXpy.

\end{enumerate}

This workflow has now been used for a number of experiments on at APS and CHESS. The complete analysis workflow can be performed in under an hour depending on the host computer. Once an initial UB matrix has been determined at one temperature, it can be optimized at other temperatures automatically, with refinements for the small changes in lattice parameter with temperature. A workflow system has been set up for data at the APS that distributes the analysis for each run to a different node on a Hadoop cluster, allowing the data reduction to be completed in the same or less time as the measurements (20 minutes per temperature). A different strategy produced similar efficiency gains at CHESS, where the CCTW operation was split into multiple segments and run in parallel in a multi-node cluster using the Swift/T data workflow language \cite{Wilde:2011jj}.

\begin{figure*}[!t]
\includegraphics[width =  2\columnwidth]{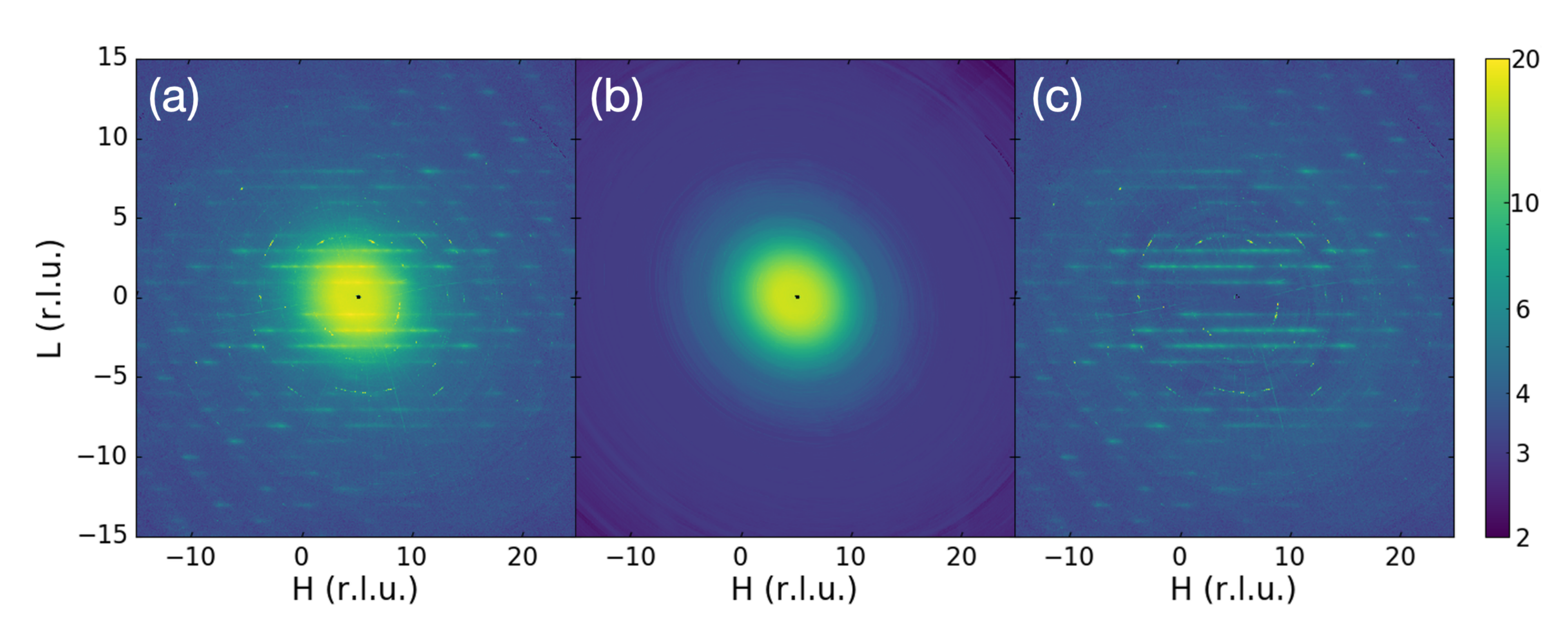}
\caption{Correction for air scattering: (a) Diffuse scattering from Na\textsubscript{0.45}V\textsubscript{2}O\textsubscript{5} in the ($H\frac{1}{2}L$) plane at 250~K. This is the original data used in Fig. 2c in the article, (b) measurement without a sample, and (c) the measurement corrected for air scattering. The intensity scale is in arbitrary units.
\label{FigureS2}}
\end{figure*}

\begin{figure}[!b]
\includegraphics[width = \columnwidth]{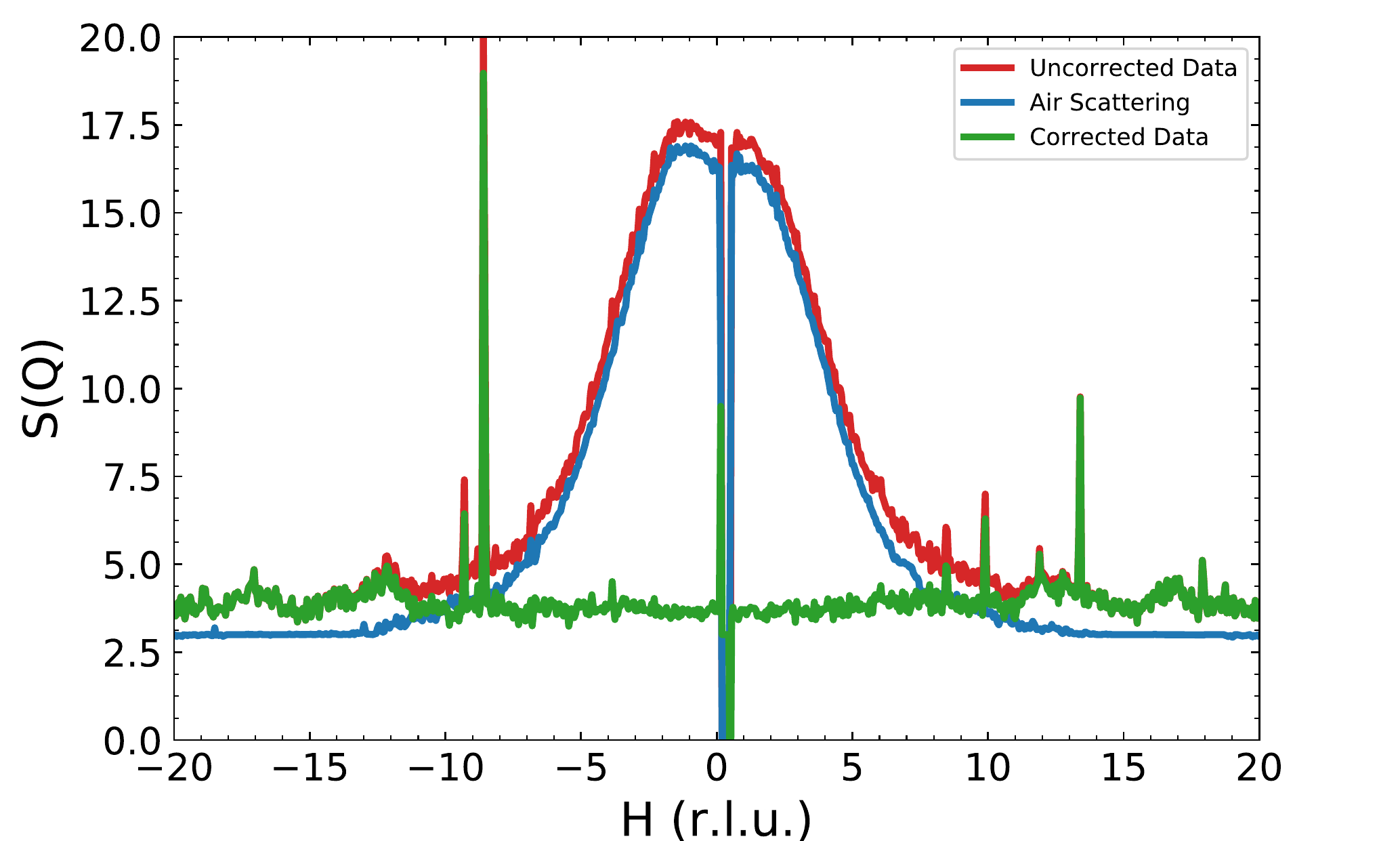}
\caption{Cuts at $L=0$ through the 2D slices shown in Fig. \ref{FigureS2}
\label{FigureS3}}
\end{figure}

\subsection{Correction for Air Scattering}
The most serious source of background in the experimental configuration shown in Fig. \ref{FigureS1} is air scattering from the incident beam between the final aperture in the incident collimation to the beam stop just behind the sample. This causes circularly symmetric scattering that is invariant during the sample rotation (Fig. \ref{FigureS2}). 

To correct for this, we performed a measurement without a sample in the beam and then used the UB matrix determined from the sample to create a three-dimensional background scan. Fig. \ref{FigureS2} shows a slice through the 250~K data at $K=0.5\pm0.02$, along with the equivalent slice in the air scattering data, and the result of subtracting the central air-scattering peak within a radius of 6.2~\AA$^{-1}$ about the origin. A cut through this data at $L=0$ is shown in Fig. \ref{FigureS3}.

Fig. 2 of the main article shows the measured data corrected for air scattering, which was used in the subsequent $\Delta$PDF analysis discussed in the next section. Even if it were not subtracted, the air scattering only adds a spherically symmetric component to the $\Delta$PDF at low distances ($<4$\AA). This would not affect the correlation length analysis shown in Fig. 3 of the main article, which is based on $\Delta$PDF values at longer distances.

\subsection{Observation of Charge Order}
\begin{figure*}[!t]
\includegraphics[width =  2\columnwidth]{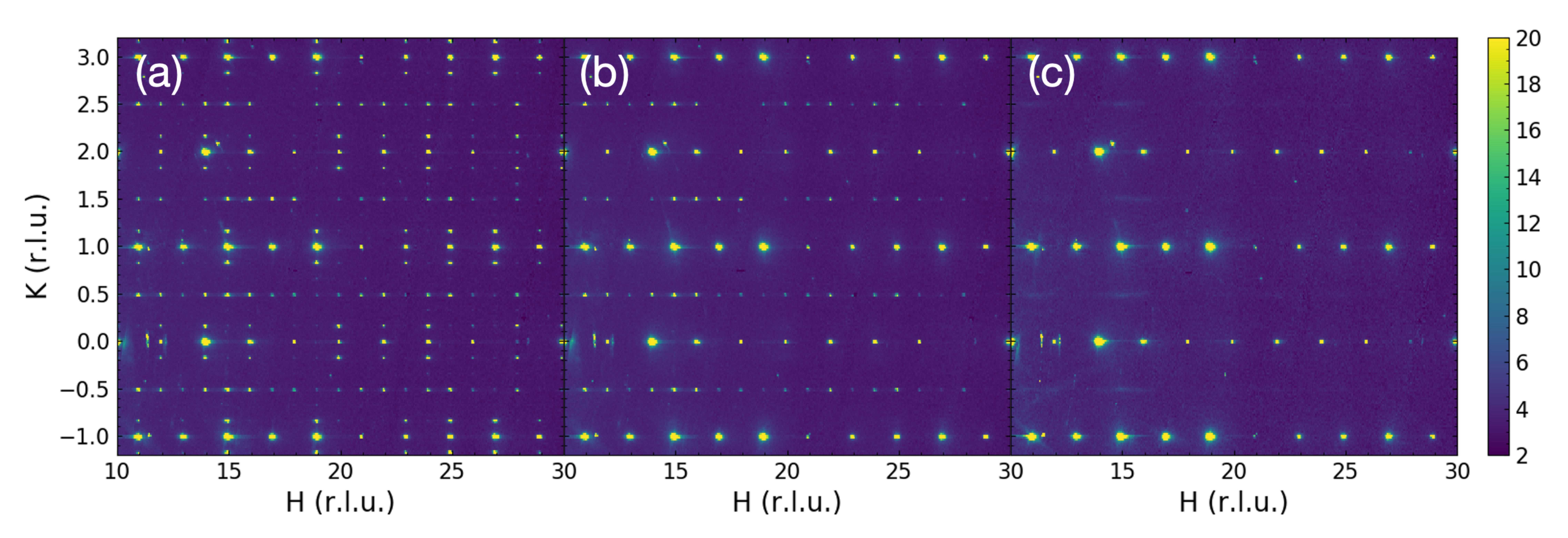}
\caption{Scattering from Na\textsubscript{0.45}V\textsubscript{2}O\textsubscript{5} in the (\textit{HK}2) plane at (a) 50~K, (b) 150~K, and (c) 250~K.
\label{FigureS4}}
\end{figure*}

Na\textsubscript{0.33}V\textsubscript{2}O\textsubscript{5} has a metal-insulator transition at 136~K \cite{Yamada:1999kn}, which has been associated with superlattice peaks observed at $K=\pm\frac{1}{6}$ with respect to the Bragg peaks of the room-temperature structure \cite{Yamaura:2002ia}. These additional peaks are evident in the data from Na\textsubscript{0.45}V\textsubscript{2}O\textsubscript{5} first appearing at $\sim$130~K. Fig. \ref{FigureS4} shows a slice through the (\textit{HK}2) plane displaying the sequence of superlattice peaks arising from structural modulations at low temperature. At 250~K, only the Bragg peaks for the $C2/m$ structure at integer (\textit{HKL}) are visible. At 150~K, the additional superlattice peaks at $K=\pm\frac{1}{2}$, which forms the main topic of this article, are evident. At 50~K, the extra peaks at $K=\pm\frac{1}{6}$ are also clear. 

\begin{figure}[!b]
\includegraphics[width = 0.7\columnwidth]{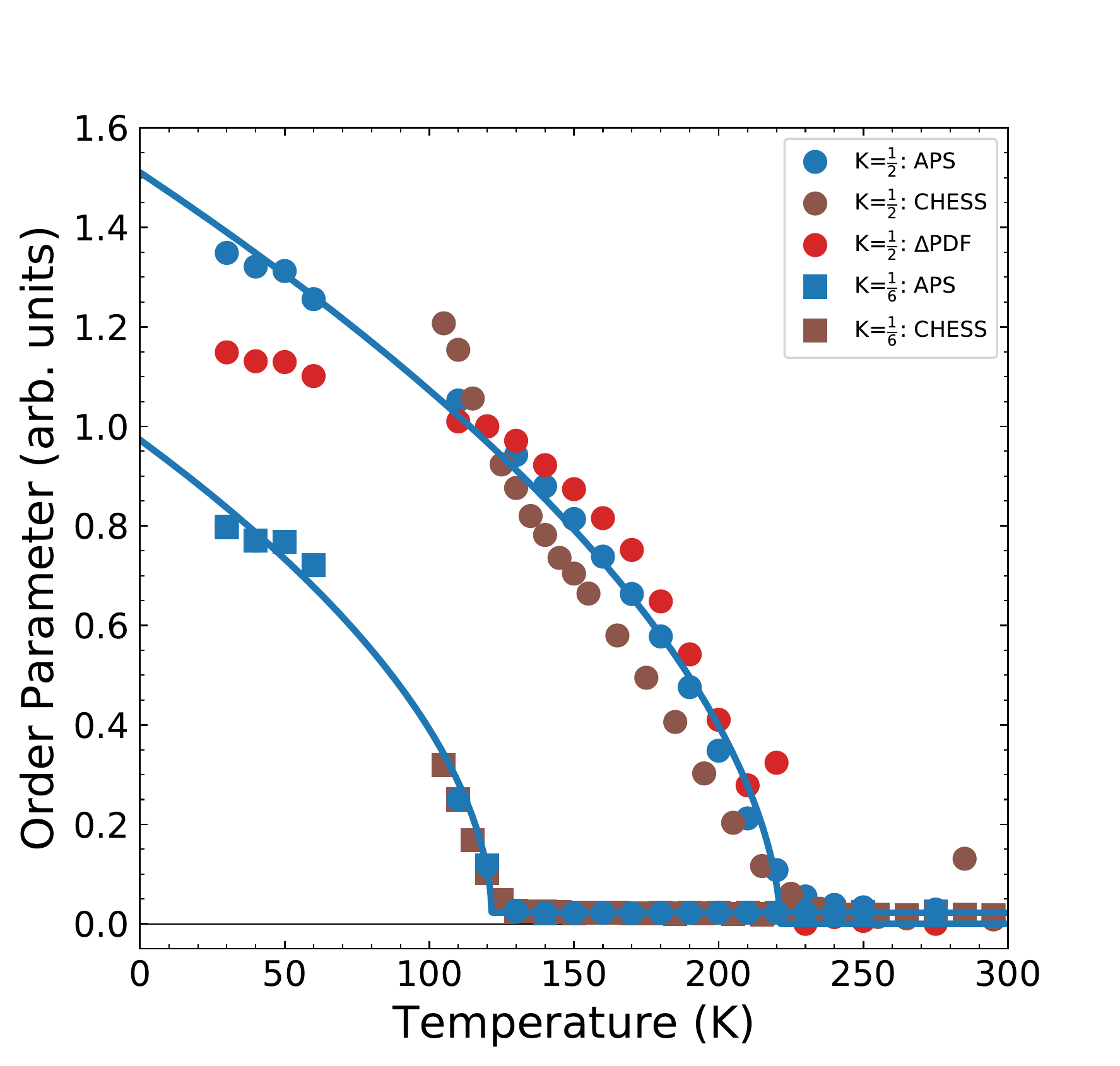}
\caption{Temperature dependence of the superlattice peaks in Na\textsubscript{0.45}V\textsubscript{2}O\textsubscript{5}. Circles (squares) represent the order parameter associated with the $K=\pm\frac{1}{2}$ ($K=\pm\frac{1}{6}$) peaks measured at APS (blue) and CHESS (brown). The red circles are the $\Delta$PDF values at $x=0$ (Fig. 4a), which approximately follow the $K=\pm\frac{1}{2}$ order parameter. The smooth lines are fits to the APS data explained in the text.
\label{FigureS5}}
\end{figure}

Although we don't believe these peaks result from long-range order, since the correlation lengths determined from the $\Delta$PDF never exceed 200\AA, their intensities do nevertheless follow a temperature dependence that resembles the order parameters of second-order phase transitions. Fig. \ref{FigureS5} shows the temperature dependence of the intensity of the superlattice peaks at $K=\pm\frac{1}{2}$ and $K=\pm\frac{1}{6}$. This was obtained by integrating all the superlattice peaks in a number of planes. We compare data measured at the APS in steps of 10~K from 30 to 250~K (apart from the missing region from 70 to 100~K caused by a beam failure) with data measured at CHESS in steps of 5~K from 100~K to 150~K and steps of 10~K from 160~K to 300~K. The temperatures of the CHESS data were shifted by +5~K to calibrate the transition temperatures to the APS data. Since the $K=\pm\frac{1}{2}$ and $K=\pm\frac{1}{6}$ integrals in Fig. \ref{FigureS5} were determined from a different number of peaks, whose structure factors vary with \textbf{Q}, the intensities are compared with arbitrary normalization. We also compare the $K=\pm\frac{1}{2}$ intensities measured directly from the total scattering with the $x=0$ value of the $\Delta$PDF shown in Fig. 4a, which has a very similar temperature dependence. 

The smooth lines in Fig. \ref{FigureS5} are the results of fits to the APS data to a power law scaling function.
\begin{equation}
I(\mathrm{T})=I(0)\left[\frac{(\mathrm{T_c-T})}{\mathrm{T_c}}\right]^{2\beta}
\end{equation}

For the $K=\pm\frac{1}{2}$ peaks, the (``pseudo-")order parameter has a T$_\mathrm{c}$ of 221.3~K with $\beta=0.285$. For the $K=\pm\frac{1}{6}$ peaks,  T$_\mathrm{c}=121.5$~K with $\beta=0.273$.

\section{PDF Analysis}

\begin{figure*}[!t]
\includegraphics[width =  2\columnwidth]{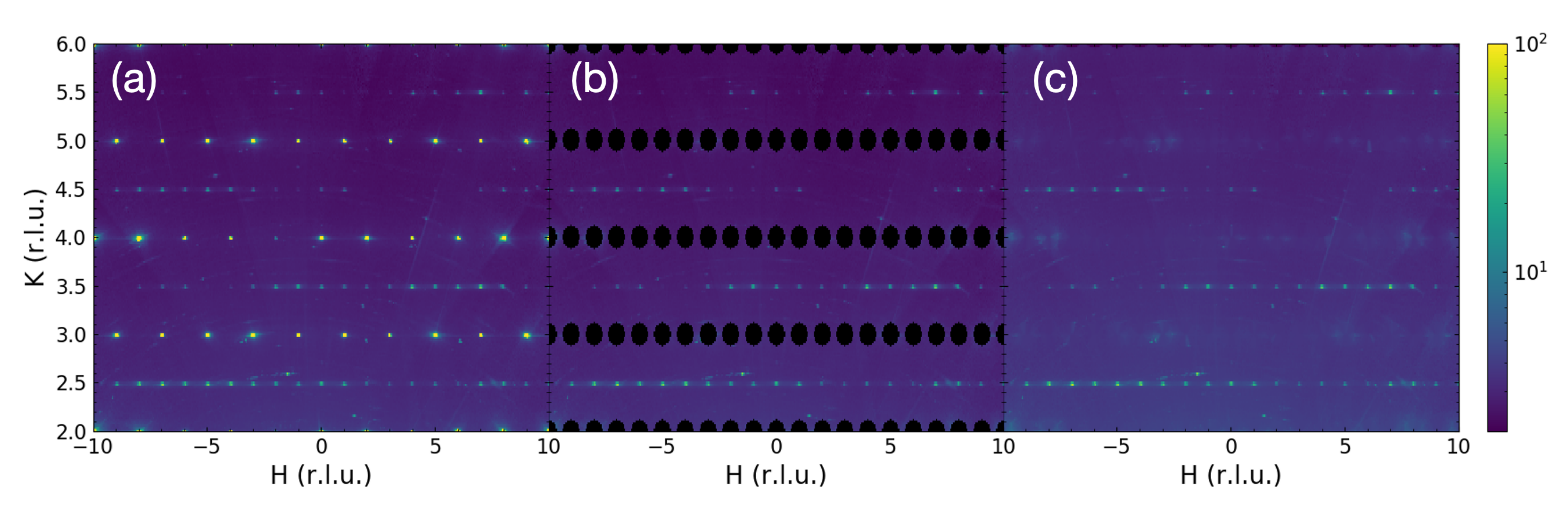}
\caption{Punch-and-Fill: (a) Symmetrized data in the (\textit{HK}0) plane, (b) the same data with a sphere of diameter 0.3~\AA$^{-1}$ around each Bragg peak at integer (\textit{HKL})s removed, and (c) the data with interpolation over the punched holes.
\label{FigureS6}}
\end{figure*}

\subsection{Punch-and-Fill Method}

We have utilized the Punch-and-Fill method developed by Thomas Weber and colleagues at ETH Z\"urich \cite{Weber:2012en, Simonov:2014vi}, in order to produce 3D-$\Delta$PDF data from the diffuse scattering. As shown in Equation 2 of the main article, removing the Bragg peaks before Fourier transforming the data produces a difference Patterson function, which only includes interatomic vector probabilities that differ from the average crystalline lattice. 

Fig. \ref{FigureS6} shows an example of the process. The three-dimensional volume containing the total scattering is first symmetrized by operations consistent with the Laue symmetry (Fig. \ref{FigureS6}a). Then a small sphere of diameter 0.3~\AA$^{-1}$ around each of the Bragg peaks at integer (\textit{HKL}) is removed (Fig. \ref{FigureS6}b). The resulting holes in the data would create further artifacts when Fourier-transformed \cite{Kobas:2005hz}, so it is essential to interpolate over them smoothly.  We used an interpolation method implemented by the Astropy convolution module \cite{astropy:2018}, which is designed to handle missing data in astrophysical observations. This requires a three-dimensional Gaussian kernel, which we developed based on the two-dimensional kernel provided by the Astropy package. The resulting data is shown in Fig. \ref{FigureS6}c.

The resulting three-dimensional volume is then transformed to real space using the Fast Fourier Transform method implemented using the Scipy fftpack module \cite{SciPy:2018}. To reduce leakage artifacts, we used the Tukey window with $\alpha=0.5$ \cite{Harris:1978ck}, which is the tapering function that best preserves the spatial resolution and minimally distorts the data.

\subsection{Analysis of Total PDF}

\begin{figure}[!b]
\includegraphics[width = 0.7\columnwidth]{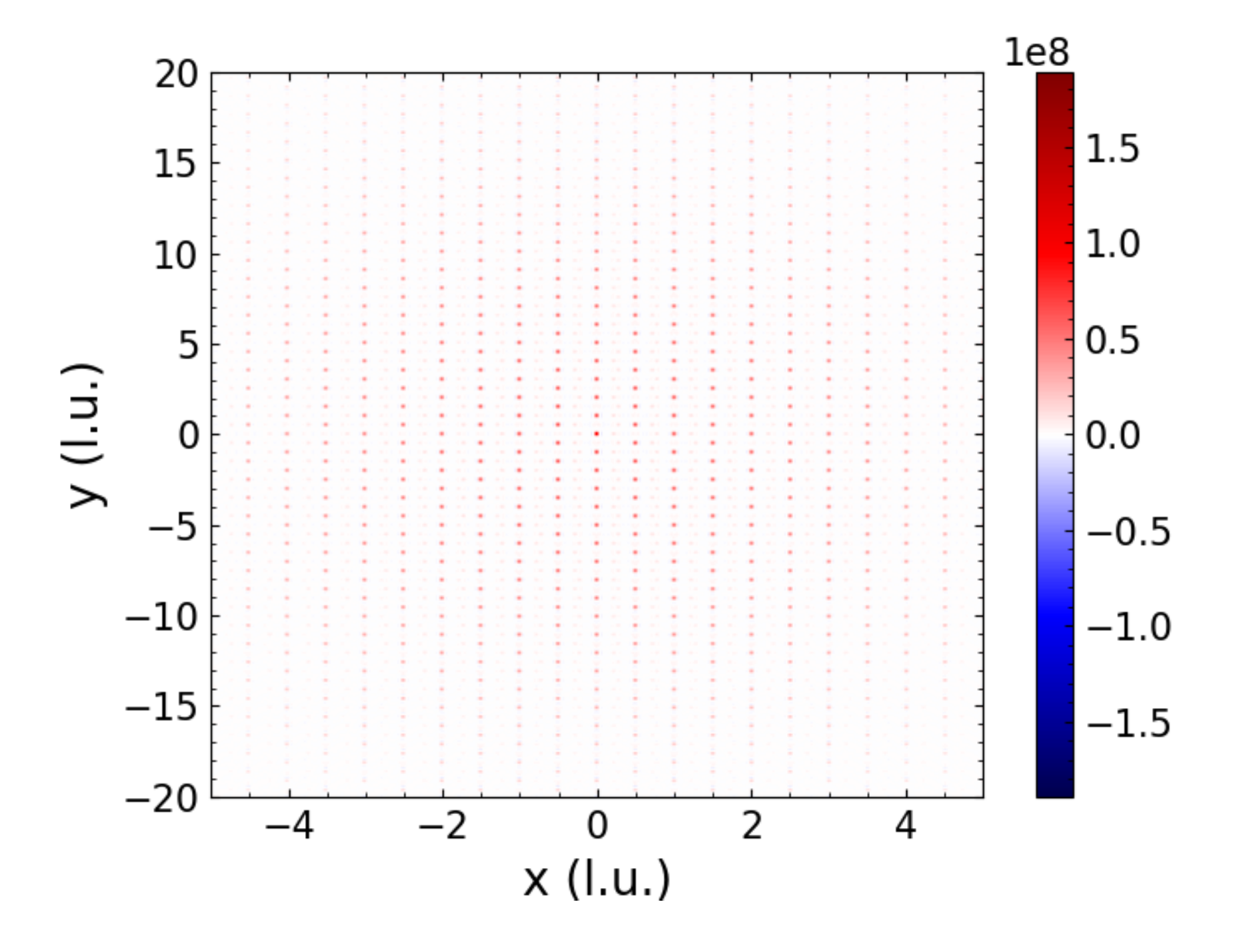}
\caption{PDF in the $x-y$ plane from Na\textsubscript{0.45}V\textsubscript{2}O\textsubscript{5} measured at 275~K
\label{FigureS7}}
\end{figure}

The scattering data has finite \textbf{Q}-resolution due to a combination of experimental parameters (beam divergence and sample mosaic) and the binning of the data. We have approximated the resolution function as a Gaussian in reciprocal space with a width that is independent of \textbf{Q}. When transformed into real space, this resolution function generates a broad Gaussian envelope that suppresses the PDF at high distance.

In order to correct for this envelope function when analyzing the $\Delta$PDF intensities, we transformed the scattering without removing the Bragg peaks in order to generate a total PDF. Since this represents the long-range structure, the PDF intensities should be independent of distance, so any modulation of the intensity will be due to the finite resolution. Fig. \ref{FigureS7} shows a $x-y$ slice at $z=0$ in the data measured at 275~K. As expected the PDF is dominated by discrete peaks at all integer values of $x$, $y$, and $z$ in lattice units (l.u. with $a=15.36$~\AA, $b=3.61$~\AA. and $c=10.05$~\AA), representing all the possible translations of the crystalline unit cell, with a monotonic decrease in intensity in all directions.

\begin{figure*}[!t]
\includegraphics[width = 2\columnwidth]{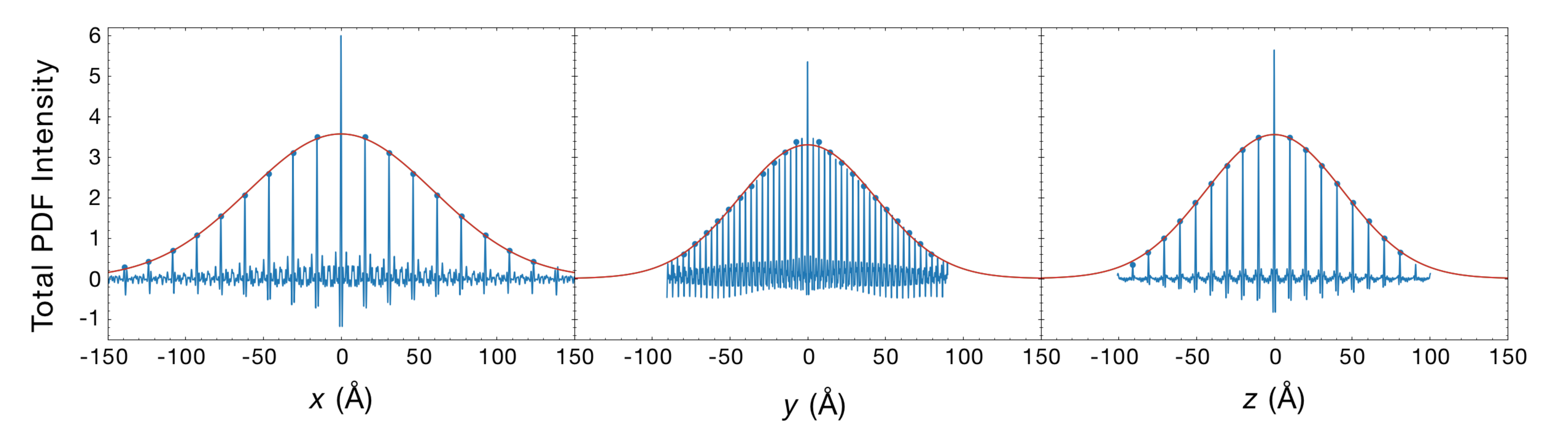}
\caption{Gaussian fits to the peak intensities of the 3D total PDF along the (a) $x$-direction, (b) $y$-direction, and (c) $z$-direction. The blue lines represent the total PDF, the blue circles are the peak values determined from the maxima, and the red line is the Gaussian fit. The extra intensity at $r=0$ arises from backgrounds that are broad in reciprocal space.
\label{FigureS8}}
\end{figure*}

To obtain the parameters of the Gaussian envelope, we have fit a Gaussian function to the peak intensities along all three crystallographic directions, excluding the $r=0$ value, which contains contributions from smoothly varying backgrounds in the measured scattering. Fig. \ref{FigureS8} shows that cuts along the three axes consist of sharp peaks at multiples of the lattice parameters with weak leakage oscillations visible between them. Systematic variations in these background oscillations along the $y$-axis may result from aliasing, but there is no evidence that this affects the analysis, since a single Gaussian gives an excellent fit over the entire range. 

The resulting standard deviations, $\sigma$, are approximately independent of temperature. 

\begin{center}
\begin{table}[!h]
\begin{tabular}{|c|c|c|c|}
\hline
\textbf{Temperature} & \textbf{$\sigma_x$ (\AA)} & \textbf{$\sigma_y$ (\AA)} & \textbf{$\sigma_z$ (\AA)} \\ \hline
30~K            & 58.93          & 40.65         & 41.97         \\ \hline
150~K           & 60.49          & 41.94         & 42.85         \\ \hline
275~K           & 59.47          & 43.69         & 44.33         \\ \hline
\end{tabular}
\end{table}
\end{center}

The bin widths of the total scattering in r.l.u. were 0.05, 0.02, 0.05, along the $H$, $K$, and $L$ directions, respectively. These correspond to $\Delta Q_H=0.020$~\AA,  $\Delta Q_K=0.035$~\AA, and $\Delta Q_L=0.031$~\AA, whose ratios approximately correspond to the ratios of their respective $\sigma$ values, so it appears that binning of the total scattering is an important component in the effective \textbf{Q}-resolution. 

\subsection{Analysis of 3D-$\Delta$PDF}

The three-dimensional arrays containing the $\Delta$PDF data generated by the Punch-and-Fill method consist almost entirely of discrete peaks occurring at $(xyz)$ positions that correspond to interatomic vectors in the \textit{C2/m} structure. The peak intensities are given by summing the probabilities of all possible atom pairs separated by the same vector weighted by their respective atomic scattering factors.  Each of these probabilities can be positive or negative if the vector occurs, respectively, more or less frequently than in the average structure. 

\begin{figure}[!b]
\includegraphics[width = \columnwidth]{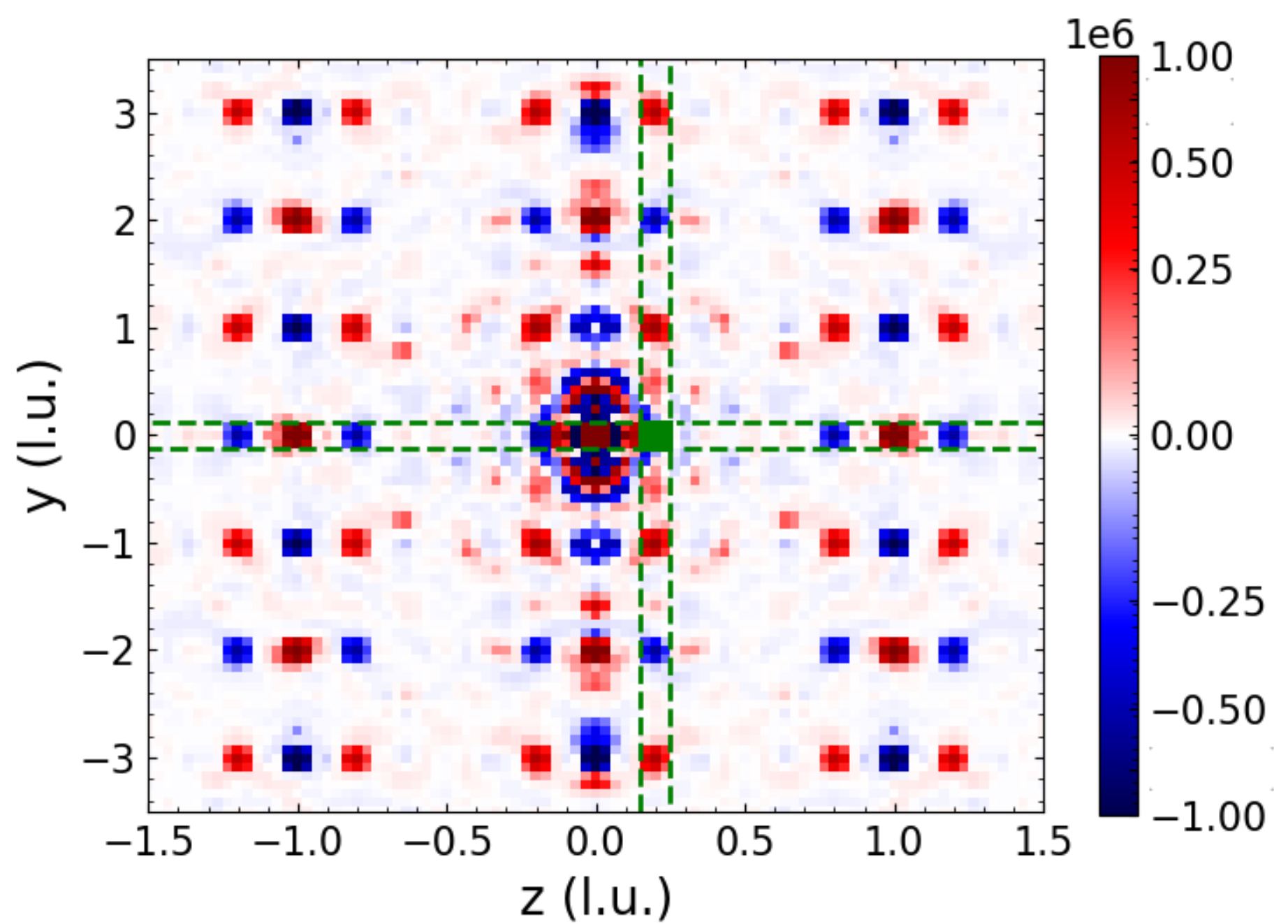}
\caption{$\Delta$PDF in the $y-z$ plane from Na\textsubscript{0.45}V\textsubscript{2}O\textsubscript{5} measured at 50~K displayed in lattice units (l.u. with $a=15.36$~\AA, $b=3.61$~\AA. and $c=10.05$~\AA). The green-dashed lines represent the cuts taken along the $y$ and $z$ axes in the subsequent analysis. The green square shows the cross section of the orthgonal cut along the $x$ axis. Peak intensities along these cuts were used in the fits shown in Fig. 3.
\label{FigureS9}}
\end{figure}

Inspection of the transformed data shows that the $\Delta$PDF in Na\textsubscript{0.45}V\textsubscript{2}O\textsubscript{5} is dominated by vectors connecting sodium ions on the network of two-leg ladders. As explained in the main article, these vectors form three-leg ladders, since the origin of the interatomic vector can be on either the left or right leg and connect to atoms on the right, same or left legs. The length of each ladder rung is $\sim$$0.18c$ or 1.8~\AA, so the triplets occur at $la+mb+nc$ and $la+mb+(n\pm0.18)c$, where $l$, $m$, and $n$ are integers (Fig. \ref{FigureS9}). These vector triplets form the strongest peaks in the $\Delta$PDF at all temperatures, confirming the conjecture that the disorder is mostly confined to the sodium sublattice in Na\textsubscript{0.45}V\textsubscript{2}O\textsubscript{5}.

Fig. \ref{FigureS9} is displayed on a symmetric log scale, in which all positive values are displayed in shades of red and all negative values are stored in shades of blue. The triplets all consist of blue-red-blue or red-blue-red spots, reflecting the stereochemical constraint preventing the simultaneous occupation of neighbouring ladder sites because the rung is so short. Furthermore, the triplets alternate between the two colour sequences showing that nearest neighbour sites along each leg are also unoccupied. Without any further modeling, it is possible to infer that the ladder sites tend to be occupied in a zig-zag pattern alternating between left and right legs from one rung to the next. 

It is also possible to infer that the zig-zag configuration in all neighbouring ladders is in phase, because our results show that probability of interatomic vectors at all values of $la+mb+nc$, are positive, provided $m$ is even. At 50~K, all the $\Delta$PDF peaks along the $x$ axis are positive, so the ladder occupations are in phase from one plane to the next. Peaks at integer values of $z$ l.u. are also positive showing that neighbouring ladders within each plane are in phase. The peaks at integer $z$ l.u. are all accompanied by negative peaks at $\sim$$\pm0.2c$. The $y$-axis peaks are positive for all even integers and negative for odd integers, because of the alternating site occupation along each leg.

\begin{figure*}[!t]
\includegraphics[width = 2\columnwidth]{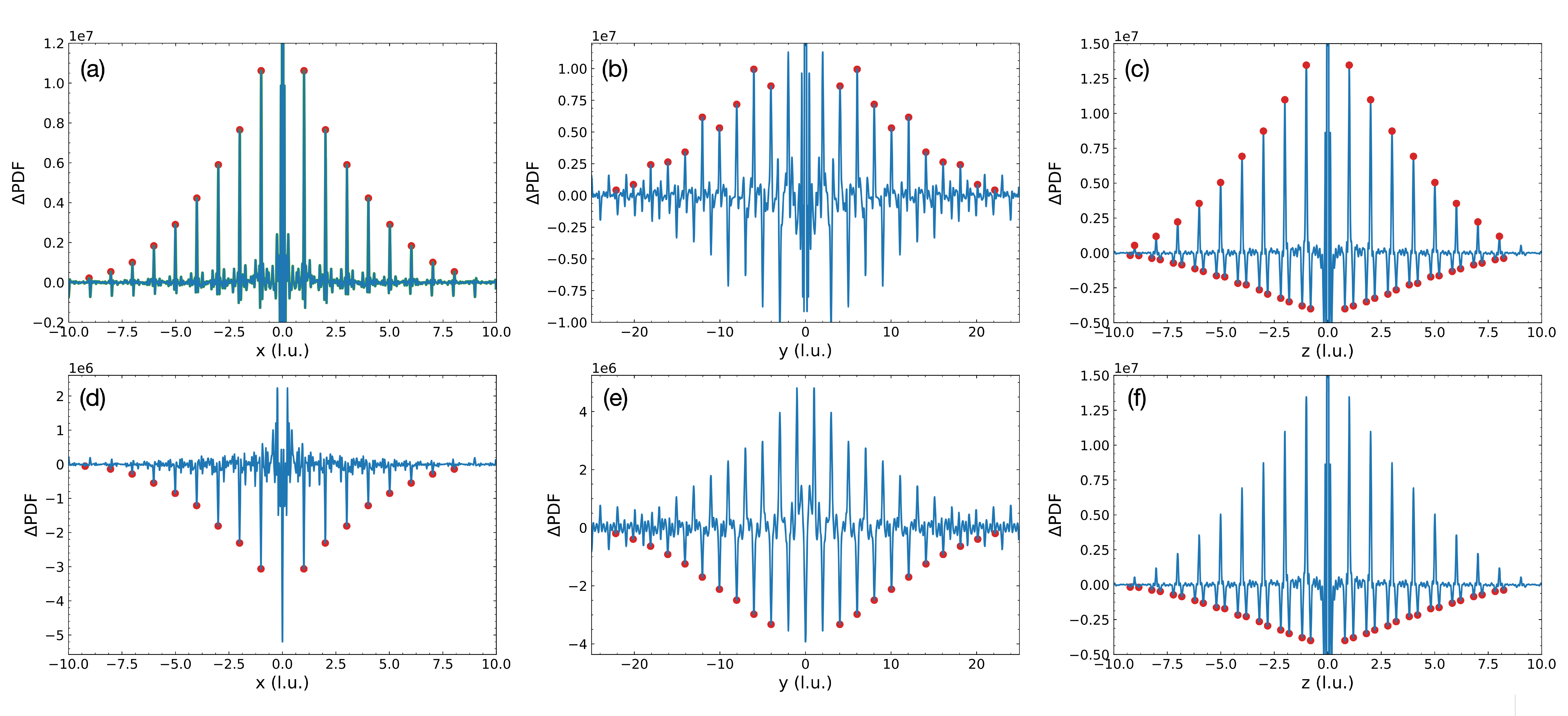}
\caption{Cuts of the $\Delta$PDF at 50~K along the three principal crystallographic directions (blue lines) along with the peak extrema (red circles). (a,b,c) Cuts that pass through the origin. (d,e,f) Average of cuts through $z=\pm0.2c$ (See Fig. \ref{FigureS9}).
\label{FigureS10}}
\end{figure*}

The size of each $\Delta$PDF peak is $\sim$3 pixels in each direction, so we have analyzed the peak intensities by taking cuts along the three axes that are three pixels wide in the two orthogonal directions (Fig. \ref{FigureS9}). Fig. \ref{FigureS10}a, b, and c shows cuts through the origin along the $x$, $y$, and $z$ axes taken at 50~K. As discussed above, all the $x$-axis and $z$-axis peaks at integer lattice units are positive, whereas the $y$-axis peaks are alternately positive and negative. However, the $y$-axis plot shows a further modulation with a period of 6 lattice units, which arises from the second phase transition at $\sim$130~K discussed in a previous section. Since integer values of $x$, $y$, and $z$ l.u. connect neighbouring unit cells, they could, in principle, contain contributions from any atoms in the structure affected by local structural distortions. The Punch-and-Fill method did not eliminate the superlattice peaks at $K=\pm\frac{1}{6}$, so the additional modulation seen along the $y$-axis could arise from modulations of the VO$_x$ polyhedra.

For this reason, we used the cuts shown in Fig. \ref{FigureS9}, which are displaced from the origin by $z=\pm0.2c$, since all such vectors are unique to the sodium sublattice. Since the peak closest to the origin is negative, all the other peaks separated by integer values of $x$ l.u. and even integers of $y$ l.u. are also negative, so we have used the negative peaks to derive the $\Delta$PDF values in our analysis of the correlation lengths shown in Fig. 3. These are shown in Fig. \ref{FigureS10}d, e , and f. Fig. \ref{FigureS10}e shows no sign of the additional $y$-axis modulation that is evident in Fig. \ref{FigureS10}b, so the sodium ladder correlations are not affected by the charge ordering, to first order.

\subsection{Exponential Decay Fits}

\begin{figure}[!b]
\begin{center}
\includegraphics[width = 0.7\columnwidth]{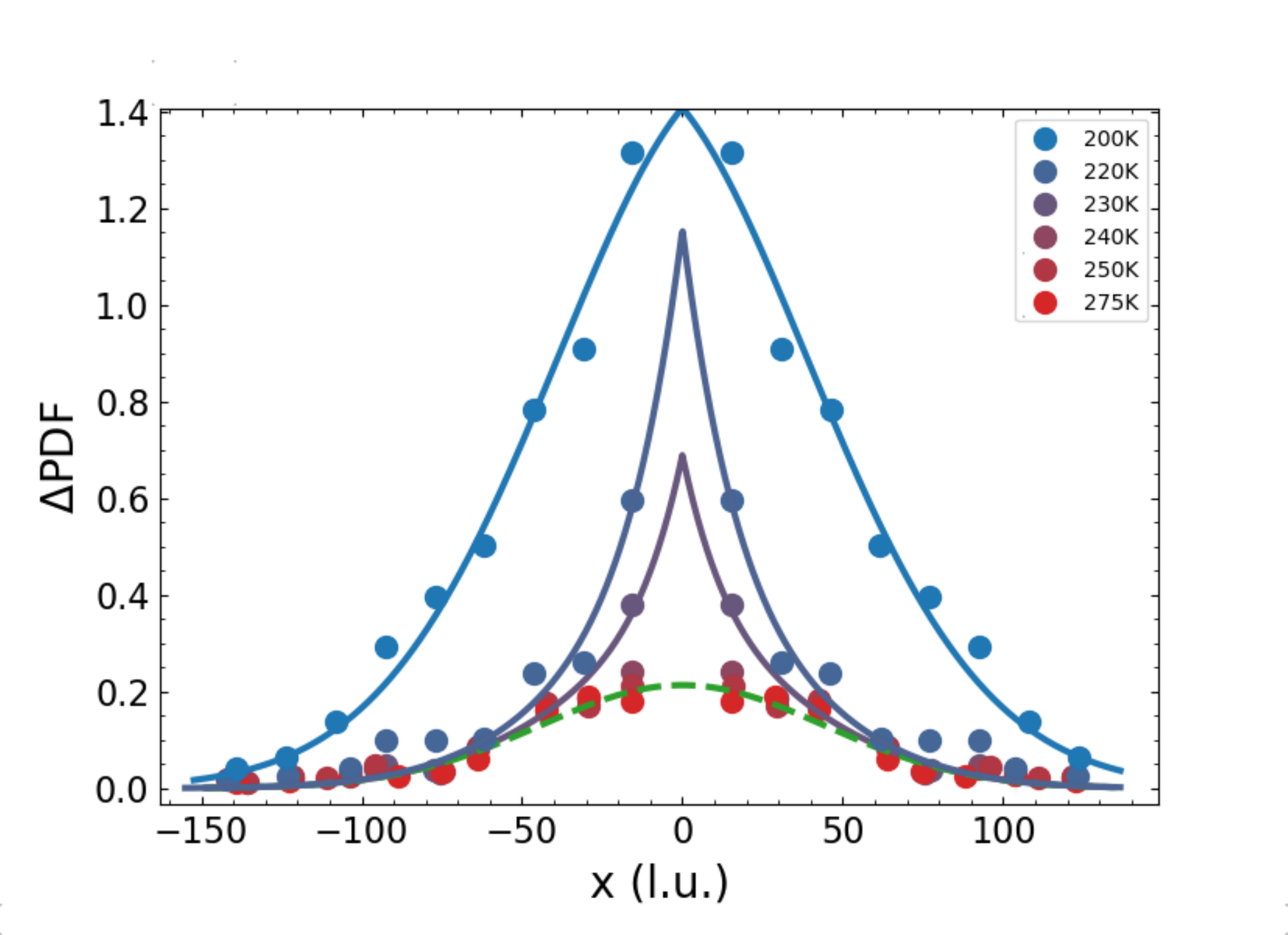}
\end{center}
\caption{$\Delta$PDF along the $x$-direction from 200~K to 275~K. A Gaussian function (green dashed line) was subtracted from the data before the fits shown in Fig. 3, which have the same intensity scale.
\label{FigureS11}}
\end{figure}

\begin{figure*}[!t]
\includegraphics[width = 2\columnwidth]{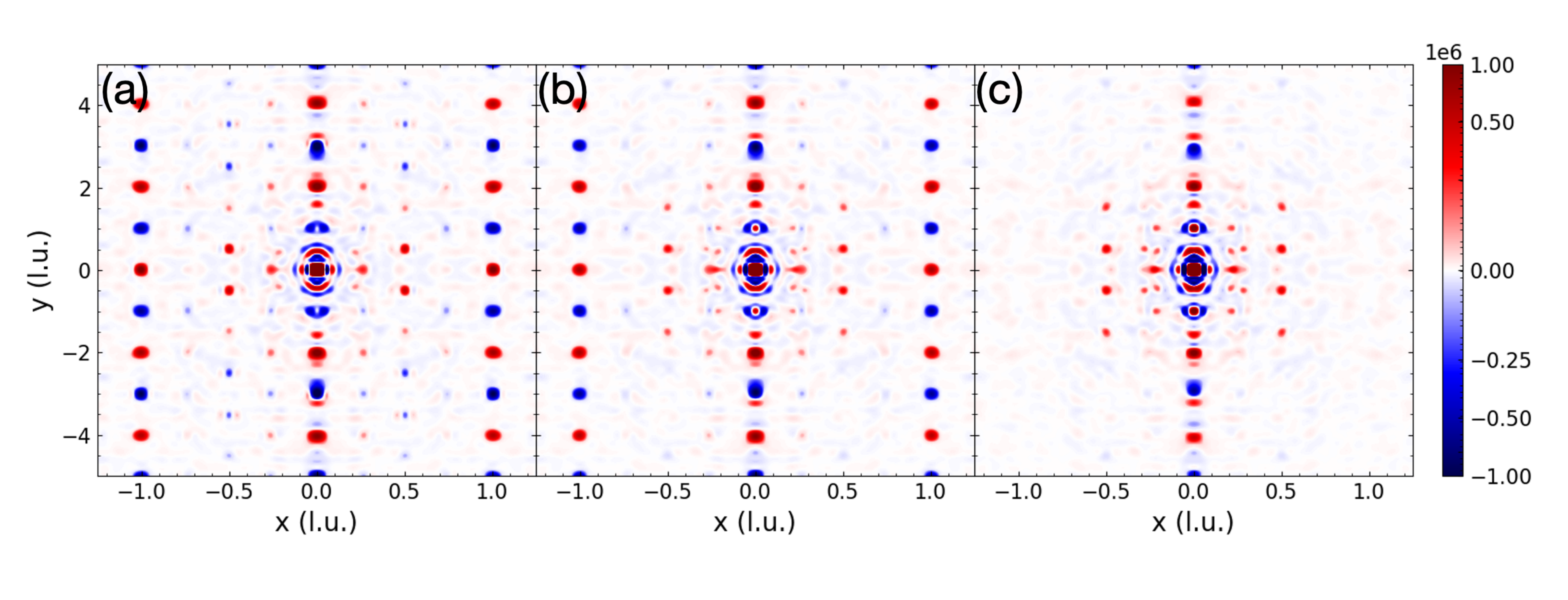}
\caption{Slice of the $\Delta$PDF in the $x-y$ plane at (a) 50~K, (b) 150~K, and (c) 250~K.
\label{FigureS12}}
\end{figure*}

The fits shown in Fig. 3 of the main article are to the product of two functions. The first is a decaying exponential, $A \mathrm{exp}\left(-r/\xi\right)$, representing the Na-Na correlations. It has two variables; $A$, the amplitude of the decay at $r=0$ in arbitrary units and $\xi$, the correlation length in Angstroms. The second is a Gaussian representing the envelope function caused by finite Q-resolution. It is centred at 0, with a standard deviation fixed to the respective $\sigma$ values for $x$, $y$, and $z$ at 275~K values shown in the table above. Along the $y$ and $z$ directions, no additional components are required to fit the data. However, along $x$, there is a weak peak centred at $x=0$ above 240~K that is inconsistent with the exponential decay. The peak is temperature-independent from 240~K to 275~K (Fig. S11). Inspection of the 3D $\Delta$PDF shows no evidence of well-defined peaks along the $x$ axies so we believe the increased PDF intensity reflects residual leakage which is known to peak in high-symmetry directions. A similar background might be present along the $y$ and $z$ axes, but it would be masked by the much stronger correlations in those directions. The green dashed line shows the background that was estimated by fitting a Gaussian function at 275~K and then subtracted from the $x$-axis data at lower temperature.

\subsection{Additional Sodium Sites}
Although the $\Delta$PDF is dominated by the triplet motifs arising from the sodium ladders, there are other weaker peaks that throw light on the disordered structure. Fig. \ref{FigureS12} shows a cut in the $x-y$ plane that reveals some of these at 50~K, 150~K, and 250~K. The peaks at $x=\pm a$ arise from the three-dimensional inter-ladder correlations that start to grow at 230~K, while those at $x=0$ arise from the persistence of correlations along the ladders above this temperature. However, there are also peaks (mostly positive) at $x=\pm\frac{1}{2}a$ that are only seen at low $y\lesssim2b$ apart from weaker negative peaks at 50~K. These occur at half-integer values of $y$ l.u., and would connect the ladders in the $x=0$ and $x=\frac{1}{2}a$ planes, since these are displaced along $y$ by $\pm\frac{1}{2}b$.

\begin{figure}[!b]
\includegraphics[width =  \columnwidth]{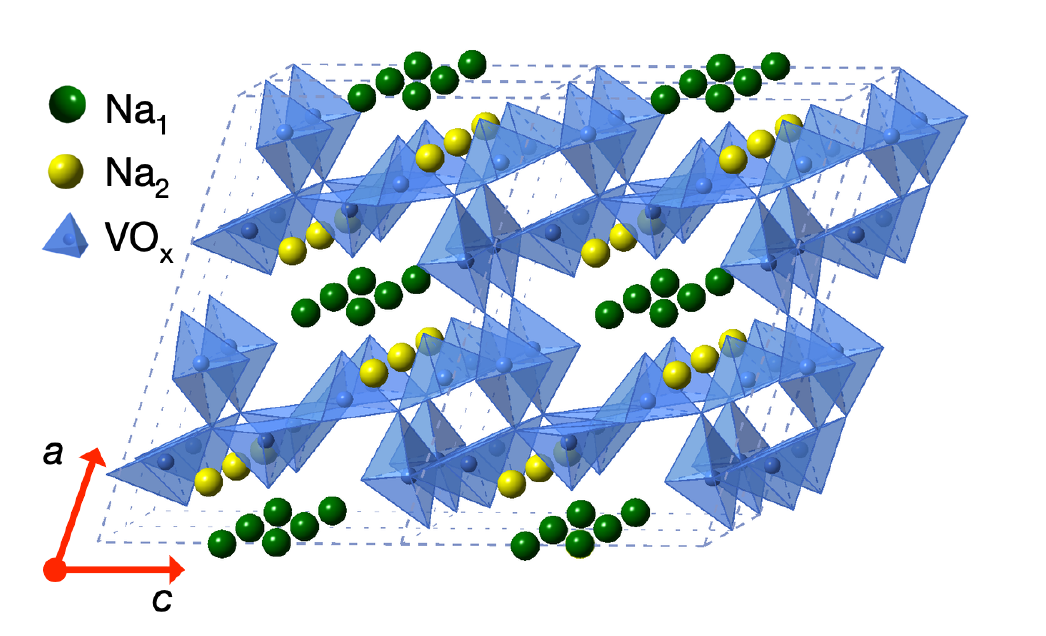}
\caption{The monoclinic structure of $\beta^\prime$-Na\textsubscript{\emph{x}}V\textsubscript{2}O\textsubscript{5} as proposed by Galy \textit{al} \cite{Galy:1970dy}. In addition to the sodium ions of the $\beta^\prime$-Na\textsubscript{\emph{x}}V\textsubscript{2}O\textsubscript{5} structure (green), they propose an additional sodium site to accommodate excess ions (yellow), which they label M$_2$. It is shown as a Na$_2$ site in the figure. 
\label{FigureS13}}
\end{figure}

It is unlikely that these $x=\frac{1}{2}a$ peaks result from correlations between the two neighbouring ladders at $x=0$ and $x=\frac{1}{2}a$. The peaks do not alternate in sign along the $y$-axis and there is no evidence of the triplet motif in the orthogonal $y-z$ plane at $x=\frac{1}{2}a$, even at 50~K (not shown). As discussed in the main article, the two possible phases of the zig-zag configuration in the $x=\frac{1}{2}a$ plane are degenerate and so should be equally probable. The resulting probability for either interatomic vector at $\frac{1}{2}a\pm\frac{1}{2}b$ should be 50\%, which is the same as in the average structure. Consequently, there should be no peak at all in the $\Delta$PDF, rather than two peaks both with increased probability. 

Another important observation is that these peaks are approximately temperature-independent, persisting to 250~K even when all the other correlations along the $x$-axis have disappeared (Fig. \ref{FigureS12}c). This implies that they have nothing to do with the site occupation within the ladders. An alternative hypothesis is that the peaks are from the excess sodium ions that occupy interstitial sites. Galy \textit{et al} proposed two possible sites \cite{Galy:1970dy}, one of which is shown in Fig. \ref{FigureS13}. This is the octahedral site that they label M$_2$, which they consider to be the most plausible to accommodate additional sodium ions. Neighbouring M$_2$ sites are indeed separated by $(\pm\frac{1}{2}\pm\frac{1}{2}0)$ lattice units, and there are no other sites in the crystal structure that share the same interatomic vectors. The limited $y$-range of the $x=\pm\frac{1}{2}a$ peaks would be consistent with the low ($\sim$18\%) occupation of these additional sites. A pinning of the phase of the zig-zag structure to these interstitial sites provides a possible mechanism for the frustration of long-range order. We therefore believe these $\Delta$PDF peaks represent strong, if perhaps not conclusive, evidence that the excess sodium ions occupy the M$_2$ sites.

\end{document}